\documentclass[preprint]{aastex}

\slugcomment{}

\shorttitle{The CFH Optical PDCS survey (COP) I: The Data}

\begin{document}

\title{The CFH Optical PDCS survey (COP) I: The Data}

\author{C. Adami}
\affil{Department of Physic and Astronomy, Northwestern University, 
Dearborn Observatory, 
2131 Sheridan, 60208-2900 Evanston, USA \\
IGRAP, Laboratoire d'Astronomie Spatiale, 
Traverse du Siphon, F-13012 Marseille, France}
\email{adami@lilith.astro.nwu.edu}

\author{B. Holden}
\affil{Department of Astronomy and Astrophysics, University of Chicago,
5640 S. Ellis Avenue, Chicago, IL 60637, USA}
\email{holden@tokyo-rose.uchicago.edu}

\author{F.J. Castander}
\affil{Observatoire Midi-Pyrenees, 14, Avenue Edouard Belin,
31400 Toulouse, France \\
Department of Astronomy and Astrophysics, University of Chicago, 5640 S.
Ellis 
Avenue, Chicago, IL 60637, USA}
\email{fjc@ast.obs-mip.fr}

\author{R.C. Nichol}
\affil{Department of Physics, Carnegie Mellon University, 5000 Forbes
Avenue,
Pittsburgh, PA 15213, USA}
\email{nichol@andrew.cmu.edu}

\author{A. Mazure}
\affil{IGRAP, Laboratoire d'Astronomie Spatiale, 
Traverse du Siphon, F-13012 Marseille, France}
\email{alain.mazure@astrsp-mrs.fr}

\author{M.P. Ulmer}
\affil{Department of Physic and Astronomy, Northwestern University, 
Dearborn Observatory, 
2131 Sheridan, 60208-2900 Evanston, USA }
\email{ulmer@curie.astro.nwu.edu}

\author{M. Postman}
\affil{STScI, 3700 San Martin Drive, Baltimore, MD 21218, USA}

\author{L. Lubin}
\affil{Palomar Observatory, California Institute of Technology, Pasadena,
CA 91125, USA}

\begin{abstract}

This paper presents and gives the COP (COP: CFHT Optical PDCS; CFHT: 
Canada-France-Hawaii 
Telescope; PDCS: Palomar Distant Cluster Survey) survey data. We describe our 
photometric and spectroscopic observations with the MOS multi-slit spectrograph 
at the CFH telescope. 

A comparison of the photometry from the PDCS (Postman et al. 1996) catalogs
and from the new images we have obtained at the CFH telescope shows that
the different magnitude systems can be cross-calibrated. After identification 
between the PDCS catalogues and our new images, we built catalogues with 
redshift, coordinates and V$_{PDCS}$, I$_{PDCS}$ and R$_{COP}$ magnitudes.
We have classified the galaxies along the lines of sight into field and
structure galaxies using a gap technique (Katgert et al. 1996). In total we 
have observed 18 significant structures along the 10 lines of sight.

\end{abstract}


\section{Introduction}

One of the main goals of the study of distant rich clusters of galaxies is
to understand their origin and evolution. Clusters are invaluable cosmological 
probes, since the evolution of cluster abundances is strongly dependent on the 
underlying cosmology and therefore can constrain cosmological models (e.g. 
Bahcall et al. 1997, Oukbir \& Blanchard 1992 and 1997, Reichart et al 1999,
Nichol et al. 1999). In order to be able to exploit this potential, large
statistically representative spectroscopic samples are, however, required. 

There are two types of samples that are being developed: those based on optical
selection criteria and those based on X-ray detections.  Below z of about 0.1,
optically selected studies (e.g. ENACS: Katgert et al. 1996) have had about as 
many clusters in  them as those selected via their X-ray fluxes. Up until 
recently, however, the higher redshift work has been dominated by X-ray 
selection techniques (e.g. CNOC: Carlberg et al. 1996 SHARC: Romer et al. 
2000; RDCS: Rosati et al. 1998; WARPS: Jones et al. 1998; Vikhlinin et al.
1998).  To greatly enlarge the sample  of detailed studies of redshift
about 0.4 optically selected clusters, we have embarked on a photometric and 
redshift campaign based on the Palomar Distant Cluster survey (Postman et al. 
1996, see also  Holden et al. 1999). 
We have observed a significant number of regions on the sky (10) and obtained 
about 70 redshifts per line of sight. These pointings were known to contain
candidate clusters of galaxies based on the PDCS studies (e.g. Postman et al. 
1996, Holden et al. 1997).

The main purpose of this paper is to publish the COP survey data and to 
describe the data reduction so as to lay a foundation for future papers. The
interpretation of the results is, therefore, given in later papers (e.g.
Holden et al. 2000). The outline of this paper is as follows. In Section 2, we 
give the observational strategy. In Section 3, we describe the way we have 
reduced and 
analyzed our photometry. In Section 4, we describe the way we have reduced and 
analyzed our spectroscopy. In the last section, we give an analysis of the 
redshift and spatial distribution of the galaxies in our sample. The data are 
given in tables 6-15.

\section{Target Selection and Observations}

\subsection{Observing strategy}

Our project required the measurement of a large number of redshifts
($\sim$ 100) of faint galaxies ($V_{PDCS}$$\leq $23) for a significant number
of clusters ($\sim$ 10). It was necessary, therefore, to optimize our
spectroscopic observations to get as many useful spectra as possible per
night. We had photometric data in our fields (the PDCS catalog: Postman et al. 
1996) prior to our CFH observations which considerably reduced the number of 
nights necessary to produce the spectroscopic catalogue (compared for example 
to the time needed to achieve the CNOC survey, Yee et al. 1996).

The first goal of the survey was to study the reality of the selected cluster
candidates: are these real physical systems or are these only
galaxy number count enhancements due to superposition effects (see Holden
et al. 2000)?  This placed a requirement on the number of redshifts we
needed along the line of sight (Katgert et al. 1996). Moreover, we wanted
to compute a global velocity dispersion for each cluster. This required
$\sim$10 redshifts in the main groups (the $clusters$) to allow us to
use robust estimators (see e.g. Adami et al. 1998c).

Assuming a line-of-sight contamination between 50 and 75\% for a cluster at
z$\sim$0.4 (see e.g. Carlberg et al. 1996) and a success rate of 70\% (see 
Adami et al. 1998b) for our magnitude ranges, we needed to obtain between 50 
and 60 spectra for each line of sight in the ideal situation of 1 cluster per 
line of sight. Since there could be 2 structures (or more) per pointing, 
however, we set a goal of measuring 100 spectra for each pointing (to yield 
about 70 redshifts).

In order to have a statistically representative set of more than 10 lines of 
sight with more than 70 redshifts, we have used the CFH-MOS multi-slits 
spectrograph for its high multiplex gain.

We wanted to measure the radial velocity of our targets with an uncertainty of
less than 150 km s$^{-1}$ because this precision is almost the same as the
one obtained for ENACS and CNOC galaxies (e.g. Katgert et al 1996, Mazure
et al 1996, Yee et al. 1996). This allows an accurate comparison with
these two surveys. Following Adami et al. (1998b) and Yee et al. (1996), we 
have used the CFH O300 grism, which provides a dispersion of about 5 \AA 
.px$^{-1}$ with the STIS2 CFH CCD (pixels of 0.43''). The precision
in the velocity measurement depends of both the resolution given by the
grism and slit width (theoretical limiting factor) and the observational
conditions (observational limiting factor). The resolution of the O300 grism
allowed us to reach the required velocity accuracy.

We observed extended objects that were a few arcsecs in diameter. In order to
properly subtract the sky in our spectra, we used slitlet lengths of 11
arcsecs. This setup allowed us to place about 40 slitlets per mask for
the full spectral range of about 6500 \AA ~delivered by the grism to be used.
The setup would require about 3 masks per pointing to reach our goal of 100,
however. In this case, the amount of time needed to observe 10 line of sights 
would have been prohibitive. To reduce the amount of time required and increase
the multiplexing capabilities of the instrument to place $\sim$70 slitlets
on each mask, we used CFH blocking filters (see Table 1 and see also the CNOC
survey: Yee et al. 1996). This was effective, since we had estimates of the 
cluster redshifts (Postman et al. 1996), which have been confirmed to be accurate (see Holden et al. 1997) enough for our purposes. These estimates 
allowed us to chose the right CFH filter so as to span a spectral range that 
included at least 3 lines for each 
spectrum (typically selected from [OII], H\&K, G band, H$\beta $, [OIII] and 
H$\alpha $). For each of the two filters used, Table 1 shows the redshift 
range that gives dispersed spectra of galaxies at that redshift that include 
the [OII], H\&K and G band spectral features.  Table 2 gives the filter used 
for each line of sight.

\subsection{Cluster candidate selection}

We selected cluster candidates to match the CFH telescope capabilities. It was 
impossible with the telescope time availability to sample structures at 
redshifts significantly greater than 0.5 (see Lubin et al. 1998 and references 
therein for such a study). We decided, therefore, to restrict our sample to 
cluster candidates in the estimated redshift range $0.3<z<0.5$.  We also 
selected clusters so as to be able to complete the sample in only two 
semesters at the CFHT. Therefore, targets were selected from the PDCS fields 
at 9 and 13 hours for the Spring semester and from the PDCS fields at 16, 0 
and 2 hours for the Autumn semester. We selected also the candidate clusters 
with both a richness class 1 or greater and significant density peaks in the 
galaxy distribution (see Fig. 1, 2 and 3) and we used the highest galaxy 
density areas. These densest areas coincided with or were close to the cluster 
centers given in Lubin et al. (1996) in most cases. PDCS34 was the exception. 
We observed at a position about 5' to the North of the given cluster center, 
slightly different from Lubin et al. (1996), as we found no galaxy 
concentration at the exact coordinates of PDCS34. 

In order to describe the galaxy distribution on the sky, we have 
computed the local projected galaxy density and produced isodensity contours 
for the PDCS galaxies using an adaptative kernel technique (e.g. Adami et al 
1998a and ref. therein) for each line of sight. This technique adapts the size
of the map window to the local density of objects. The same technique has been
used in Adami et al (1998a) to study the ENACS clusters. Where the galaxy 
density is higher, the window used to compute the density of objects is reduced 
to a value consistent with producing a statistically significant number of 
galaxies. Where the galaxy density is lower, the window is larger so as to 
produce a similarly valid statistic for the density estimate. We have then 
produced the Figures 1,2 and 3.

Finally, among the cluster candidates matching the previous conditions, we 
selected those ones that were detected in X-rays (Holden et al. 1997, 1999)
whenever possible (only 3 of the 10 lines of sight).
 
\subsection{Mask design}

We optimized the number of slitlets per mask to increase the efficiency of the
survey. It is possible to show that for a field with a very high density of 
targets, the optimal configuration is to place the slitlets in band 
configurations. With our limiting magnitude and slitlet width, however, this is
not the best method to use because the density of targets is not always the 
same. To take this into account, we have adapted the Minimal Spanning Tree 
(MST hereafter) method (e.g. Dussert et al. 1986) in order to find the optimal 
configuration according to the filter used. We only give a brief description
here: for a given set of 
points, the MST process finds the minimal total length of a tree covering this 
set (without a loop). If we fix the area (thus the length of tree) the MST 
exactly finds the maximum number of slits that can be put in that area 
(according to the constraints: size of the slits, magnitudes, filter ..etc...).
We have checked this method by showing that it gives a band configuration for a
high density of targets. We show in Figures 4 and 5 the typical configuration 
given by this method for 2 of the lines of sight: PDCS62 and PDCS67. PDCS62 has
a high density of targets (3.97 gal arcmin$^2$) while PDCS67 has a lower target
density (2.20 gal arcmin$^2$). The slit distribution for PDCS62 is close to a 
band configuration. 

Practically, we designed the masks in three steps. First, the primary potential
targets were selected from the galaxies with a low enough magnitude to provide 
a reasonable success rate (percentage of observed galaxies with a redshift 
successfully measured) according to the planned exposure time. This exposure 
time was chosen to observe galaxies brighter than V$_{PDCS} \sim -19$ at the
mean redshift of the survey (z$\sim$0.4). This is about 1.5 magnitude fainter
than the typical values of M$^{*}$ in nearby clusters (see e.g. Rauzy et al. 
1998).

We used the MST selection for these galaxies first. Table 2 gives these 
magnitude ranges with the real success rates. The mean value is 66$\%$, only 
slightly lower than the expected value. Then, the 
secondary potential targets were selected from the galaxies in the next 0.5 
magnitude bin (in principle too faint to provide the same S/N, see Table 2 for 
the $V_{PDCS}$ magnitude range). Finally, if some space remained on the mask 
after selecting these two types of targets, we also assigned slitlets to 
contain other objects, typically in the $V_{PDCS}$ magnitude range [22,23] 
or selected by "eye" during the night (tertiary targets).

We did not select the galaxies on the basis of their color, in order to avoid 
selection effects along the line of sight. Also, the second mask for PDCS62 
has been partially designed by hand (during the night with the image acquired 
at the CFH telescope) because the PDCS photometric data did not cover all the 
field (only about 50$\%$).

\section{The photometry}

\subsection{The PDCS data}

We selected 10 lines of sight (see Table 2 and Fig. 1-3 and 6) for our
observations which include 14 PDCS cluster candidates that are described
in Postman et al. (1996).  We have used the original PDCS photometry to
select the galaxy targets. The PDCS photometry was carried out in the 4-shooter 
Palomar $V$ and $I$ filters and calibrated in the AB system. We will refer to 
it as $V_{PDCS}$ and $I_{PDCS}$ from now on. Postman et al. (1996) showed 
how this photometry compared to the Vega-system standard system. Here, we 
briefly describe the comparison between systems: the effective wavelength of
the $V_{PDCS}$ filter is $\simeq$100\AA ~bluer and about 50$\%$ wider than 
the standard Johnson's $V$. $I_{PDCS}$ has nearly the same width as the 
Kron-Cousins $I$ filter, but the effective wavelength is about 500\AA  ~redder.
The zero points of the $V_{PDCS}$ and $I_{PDCS}$ magnitudes are based on the 
$AB$ magnitude system of Oke $\&$ Gunn (1983). The magnitudes of Vega are 
$V_{PDCS}$=+0.03 and $I_{PDCS}$=+0.46.  The relation between ($V_{PDCS}$, 
$I_{PDCS}$) and (V,I) are:

$V = V_{PDCS} -0.02-0.056(V_{PDCS} -I_{PDCS} )+0.012(V_{PDCS} -I_{PDCS})^2$

and

$I = I_{PDCS} -0.43+0.089(V_{PDCS} -I_{PDCS})$

The uncertainty for $V_{PDCS} - I_{PDCS}$ is almost 0.2 magnitude (Lubin 1996).

Since we used the PDCS star/galaxy classification to select the galaxies to
observe, it is of interest to determine how well this selection performed. 
Given our observational strategy with blocking filters, faint
stars remain unidentified because no obvious spectral feature falls within
our spectral coverage. The same is true for faint galaxies at redshifts
outside the optimal range of the filter used and for galaxies only
detected at low signal-to-noise. The validity of the star/galaxy separation can
only then be tested against other methods for these cases.  We have thus
compared the PDCS star/galaxy selection against the classification scheme
of Sextractor (Bertin $\&$ Arnouts 1996). Figure 7 shows the
comparison for the West PDCS62 spectroscopic field using a $V$ image taken
at CFH (see below). We have plotted the ANN (Artificial Neural Network)
parameter of Sextractor characterizing the nature of the objects versus
the $V_{PDCS}$ magnitude for the PDCS objects classified as $galaxies$. The 
objects in Figure 7 are, therefore, only galaxies according to the PDCS
classification. The Sextractor ANN parameter spans the range [0,1], being 
close to 1 if the object is classified by Sextractor as a star and moving 
closer to 0 if the 
object resembles a galaxy. There is a low contamination rate at faint 
magnitudes: a few number of objects classified as galaxies by the PDCS are
interpreted as stars if we use Sextractor. We conclude, therefore, that we 
optimized the selection of our targets as can be seen by the "ridge line" near 
0 in Figure 7. However, according to the mask design technique, we targeted
sometimes, in the tertiary-target-class, objects that were classified as stars 
by the PDCS, but which were later found to be galaxies. These objects 
represent, however, less than 3$\%$ of the sample and are not a significant 
source of error.

\subsection{The CFH photometry}

Besides the multislit spectroscopy, we have also imaged the fields of
study with the CFHT. The field of view of the frames obtained was $10'
\times 10'$. The imaged areas are shown in Fig. 6 (the photometric
fields are slightly larger than the spectroscopic fields of Fig. 6 which
cover about $8' \times 8'$).  We used the $R$, $V$, $2503$ and $4611$
filters. The $2503$ and $4611$ filters were the blocking filters used for
the spectroscopy (to limit the extent of the spectra) and are described in
Table 1. They are, respectively, a very wide $V$ filter and a filter
similar to a combination of the $V$+$R$ filters. The $V$ filter is a
standard Johnson filter (centered at 5470\AA , and FWHM of 880\AA ) and the
$R$ filter is similar to the Kron-Cousins $R$ but somewhat narrower
without the red tail. It is centered at 6500\AA , and has a FWHM of 1280\AA . 
From now on we will refer to these filters as $V_{COP}$ and $R_{COP}$.

All fields were imaged for 5 minutes, except for PDCS16 and the two first
fields of PDCS38 that were exposed for 15 and 8 minutes, respectively (see 
Table 3). The images have been photometrically calibrated in the Vega system 
using several Landolt standard fields (Landolt 1992). Given that most of the
fields were observed in only one filter, the photometric calibrations only
include a extinction term and a zero point but not a color term. The
uncertainties in our photometry were dominated by the fluctuations of the zero
points, computed at different airmasses throughout the night, due to
imperfect photometric conditions. The internal statistical errors
within a field are negligible, except for the faintest objects. We estimate 
the systematic zero-point error in our measured magnitudes to be less than 
0.15 mag for the observations taken in February 1998 (see Table 3) and less 
than 0.10 mag for the August 1998 observations. Table 3 summarizes our 
observations.

We used the $V_{COP}$ exposure for the second field of PDCS62 to complete the 
$V_{PDCS}$ data. This cluster was not completely covered by the PDCS 
photometry. To transform our magnitudes into a $V_{PDCS}$, we have computed 
the relation between $V_{PDCS}$ and our $V_{COP}$ for the first field of 
PDCS62 and applied it to the second field where the $V_{COP}$ magnitudes were
also available. The best fit obtained was:

$V_{PDCS} - V_{COP} = -0.28 (V_{COP} - block_{4611}) + 0.14$

We plot finally on Fig. 8 the relations between $V_{PDCS}$ and $I_{PDCS}$,
and between $V_{PDCS}$ and $R_{COP}$ for all the galaxies in our sample. 

\subsection{Galactic extinction}

The fields chosen for the PDCS were selected from high-latitude Gun \& Oke
(1975) survey areas. We avoided regions of high extinction. As expected, the
galactic extinction values obtained from the Burstein \& Heiles (1982) and
Schlegel et al (1998) reddening maps are low. The mean extinction in the
$V_{PDCS}$ band is 0.027 and lower than this value for the $I_{PDCS}$ filter 
in all the fields (Postman et al 1996). Given the small extinction values and 
our photometric errors we have chosen not to correct for extinction.

\section{The CFH spectroscopy}

\subsection{Computing the redshifts}

We have used both the MIDAS (public ESO reduction package) and IRAF (see e.g. 
Kurtz \& Mink 1998) packages to reduce the 2-Dimensional spectra to 
1-Dimensional spectra. The details of the method used can be found in Holden 
et al. (1999). We computed the redshift from the 1-dimensional spectra from 
emission lines and from cross-correlation techniques (e.g. Tonry $\&$ Davis 
1979). If there were more than two emission lines ("only"), we computed 
the emission line redshift measuring the centroid of the identified lines using 
gaussian fits and averaging the redshifts. For absorption line dominated 
spectra, we cross-correlated the spectra with 4 different spectroscopic 
templates (M31, M32, a 20 Gyear old E/SO Bruzual \& Charlot (1993) model and 
finally a spectrum resulting from the combination of 1959 low-z high quality 
absorption line spectra: Kurtz \& Mink 1998). We used the IRAF/RVSAO package to 
compute the redshifts. For absorption line dominated galaxies, we produced 4 
estimates of the redshift, one from each template. To select a unique value, 
we proceeded as follows:

-1st: eliminated all the redshift estimates lower than -0.015 assuming
that even the infalling galaxies of the Virgo cluster have velocities greater
than -4500 km s$^{-1}$. This limit is clearly an extreme value.

-2nd: eliminated all the redshift estimates with a cross-correlation
coefficient lower than 3 (see Kurtz $\&$ Mink 1998, Tonry $\&$ Davis 1979).

-3rd: selected as the true redshift the estimate with the best
cross-correlation coefficient if there was a gap of more than 1 with the
second best coefficient.

-4th: assumed the mean value of the redshifts with the best
cross-correlation coefficients if they were in agreement (difference less
than 300 km s$^{-1}$).

-5th: if neither the 3rd nor 4th conditions were fulfilled, we have simply
assumed the value of the redshift with the best cross-correlation coefficient.

Using this approach, we obtained 636 redshifts (see Table 2 and 4 for details).
We present a sub-sample of 4 spectra in Fig 9. The lower left spectrum is 
typical of our best signal to noise. It represents a galaxy at z=0.462
with a cross-correlation coefficient of 13.26. The lower right spectrum is
a galaxy where we have used emission lines to deduce the redshift (z=0.658), 
e.g. the [OII] line shown on the figure (at $\sim$6180\AA). This spectrum is 
typical of the worse spectra we used, but the cross-correlation method also was 
able to detect the CaII H$\&$K lines around 6600\AA. The upper spectra (left 
and right) are typical of all our sample. The upper left spectrum is a galaxy 
at z=0.461 (cross-correlation coefficient of 4.33) and the upper right spectrum 
is a galaxy at z=0.459 with both absorption line features (cross-correlation 
coefficient of 3.54) and emission line features ([OII]).

\subsection{Checking the redshifts}

To check the validity of the assigned redshifts we have "eye-balled" all the
emission line redshifts and also checked a randomly selected sample of
absorption line dominated spectra. Our visual inspection confirmed the validity 
of our computationally derived redshifts.

For 19 objects, we also had two separate spectra and, hence, two independent
measurements of the redshift. These objects were observed twice due to
overlaps in observing runs done at CFH and at the 4 meter Mayall telescope (see 
Holden et al. 1997). In order to estimate the uncertainties of our redshifts,
we have then plotted the percentage difference between the two estimates
versus the cross-correlation coefficient $r$ (Tonry $\&$ Davis 1979). Fig. 10 
shows the 18 galaxies with less than 2.5$\%$ of difference. The mean error for 
the redshift estimate is 0.7\% of the redshift (or 0.0016 in redshift).  The 
19$^{th}$ galaxy, although having a cross-correlation coefficient of $r$=4.2, 
had a discrepancy of 42\% . This is clearly due to one wrong redshift. This 
galaxy was observed twice at CFHT. The first observation provides a redshift of
0.15364 and a correlation coefficient of 4.2. The second observation yielded a
redshift of 0.08958 and a correlation coefficient of 4.2. However, for this
observation, the second best estimate (with another template) of the redshift 
is 0.15673 with a correlation coefficient of 3.71, still acceptable. Using the
second value of the redshift, the difference is only 2$\%$ for the initially 
discrepant galaxy. Assuming that these 19 galaxies are representative of our 
entire sample, we estimate that less than about 5\% of our sample has a false 
redshift assignment. This has a negligible effect on conclusions we draw
based on these results.

\section{Analysis}

\subsection{Final catalogues}

We identified the objects we measured at CFH (redshift + R$_{COP}$ magnitude) 
with the galaxies in the PDCS survey in order to build catalogues with 
position (measured at CFH), redshift, R$_{COP}$ magnitude and V$_{PDCS}$ and 
I$_{PDCS}$ magnitude (Tables 6-15). This is also a way to estimate the 
uncertainty 
for the coordinates of the galaxies. We found a mean difference between the 
PDCS coordinates and the coordinates measured at CFH of 3.5''$\pm$2.3''. This 
is typically the uncertainty for the coordinates we give in table 6-15.

We also classified the galaxies in redshift space as members of a structure or 
as field galaxies. This was a first step. A more detailed classification is 
discussed in Holden et al. (2000), but we give these results in order to 
present a complete overview of the data. In order to make this classification, 
we have searched the velocity distribution of each line of sight for gaps of 
more than 1000 km s$^{-1}$. If we had more than 5 galaxies between two 
successive gaps, we have called these galaxies a {\it structure}. This is 
exactly the same method used to define the structures in the ENACS catalog 
(Katgert et al. 1996). The method does not completely avoid the inclusion of 
some interloppers, but, to a first approximation, it defines the compact 
structures (gravitationally bound) in redshift space. We summarize the 
results in Table 4, Table 6 to 15 and in Figures 11 and 12.

\subsection{Completeness and spatial representativity}

We show in Fig. 13 the variation of the completeness level of the spectroscopic
catalogue compared to the photometric catalogue. This completeness level $C$ is
defined as the ratio between the number of galaxies with a measured redshift
(galaxies targeted $and$ successfully measured) and the total number of 
galaxies. It is different from the success rate, which is only the ability to 
deduce the redshift of a target. We see on Fig. 13 that this 
level is constant around 35$\%$ from $V_{PDCS}$=16.5 to $V_{PDCS}$=21.0. This 
relatively low level is because we did not have time to put a slit on all the 
available galaxies. The percentages of the 2 brightest bins of 
Fig. 13 are based on a low number of galaxies, since, we targeted more faint 
galaxies than bright galaxies explicitely to try to keep constant the 
completeness level $C$. This constant sampling is important for studying the 
galaxy distribution along the line of sights because it prevents us from severe 
redshift selection effects. For the faintest galaxies, the completeness level 
drops down to 6$\%$ for $V_{PDCS}$=22.5. These percentages were computed using 
all the lines of sight put together except PDCS61 for which the exposure time 
was very short. These percentages do not change considerably from pointing to 
pointing (except for PDCS61). For several types
of analysis, it may be useful to give an analytical expression of the
variation of the completeness level $C$. For the magnitudes brighter than
$V_{PDCS}$=20.5, $C$=35.5$\%$. For the fainter magnitudes, assuming a
power law model, the best fit is:

$C = 10^{-0.44(V_{PDCS} - 24.07)} \%$

To test for selection effects in the spatial distribution of the 
galaxies for which we measured redshift, we compared the spatial distribution
of our redshift measured sample to that of all the PDCS galaxies. For this
comparison, we have used a bidimensional Kolmogorov-Smirnov test as 
a function of the V limiting magnitude to determine the variation of this 
representativity level as a function of the photometric depth of the sample. 
A value of the representativity level given by the Kolmogorov-Smirnov test 
close to 100$\%$ means that the two distributions on the sky are very similar:
the galaxies with a measured redshift are a statistically representative 
sub-sample in term of spatial distribution. A value lower than 90$\%$ means 
that this sub-sample is statistically different at the level of 10$\%$. We see 
in Tab. 5 that, except for PDCS30-45, the two spatial distributions are 
indistinguishable for the magnitudes brighter than 21. Note that the case of 
PDCS30-45 is not easely explained (see Tab. 5).

\section{Summary}
 
We have presented and given the data gathered in the COP survey. The 
spectroscopic and 
photometric observations were performed with the MOS/STIS2 instrument during 6 nights at the CFH telescope with the grism O300 and 2 blocking filters to
enhance the multiplex gain of MOS. We have used a method based on the MST 
theory to optimize the number of slits per mask. This allowed us to measure 
636 redshifts for 10 PDCS lines of sight. These lines of sight were selected
to hold PDCS candidate clusters, with significant peaks in the galaxy density
distribution.

The success rate (percentage of targeted galaxies with a successfully measured
redshift) was close to 70$\%$ for the primary targets (typically brighter than
V$_{PDCS}$=22.). The completeness level (percentage of all galaxies with a
measured redshift) was about 35$\%$ down to V$_{PDCS}$=20.5. The galaxies with
a redshift were proved to be a spatially representative sub-sample down to 
V$_{PDCS}$=20.5 (no significant spatial selection effects). Finally, the 
percentage of false redshifts was about 5$\%$, based on 19 galaxies observed 
twice.

A comparison of the photometry from the PDCS (Postman et al. 1996) catalogs
and from the new images we have obtained at the CFH telescope shows that
the different magnitude systems can be cross-calibrated. This confirmation is
important for the reliability of future works based on the multi-color
photometry of COP. After identification between the PDCS catalogues and our
new images, we built catalogues with redshift, coordinates and V$_{PDCS}$,
I$_{PDCS}$ and R$_{COP}$ magnitude (Tab. 6-15).

We have classified the galaxies along the lines of sight into field and
structure galaxies using a gap technique (Katgert et al. 1996). In total we 
have observed 18 significant structures along the 10 lines of sight (Tab. 4).
As noted in the introduction, the interpretation of the results is given 
elsewhere (e.g. Holden et al. 2000).

\acknowledgments

CA thanks the staff of the Dearborn Observatory for their hospitality 
during his postdoctoral fellowship. The authors thanks the CFHT TAC for
support. BH would like to acknowledge support from the following: NSF
AST-9256606, NASA grant NAG5-3202, NASA GO-06838.01-95A, and the
Center for Astrophysical Research in Antarctica, a National Science
Foundation Science and Technology Center.

\clearpage

\figcaption[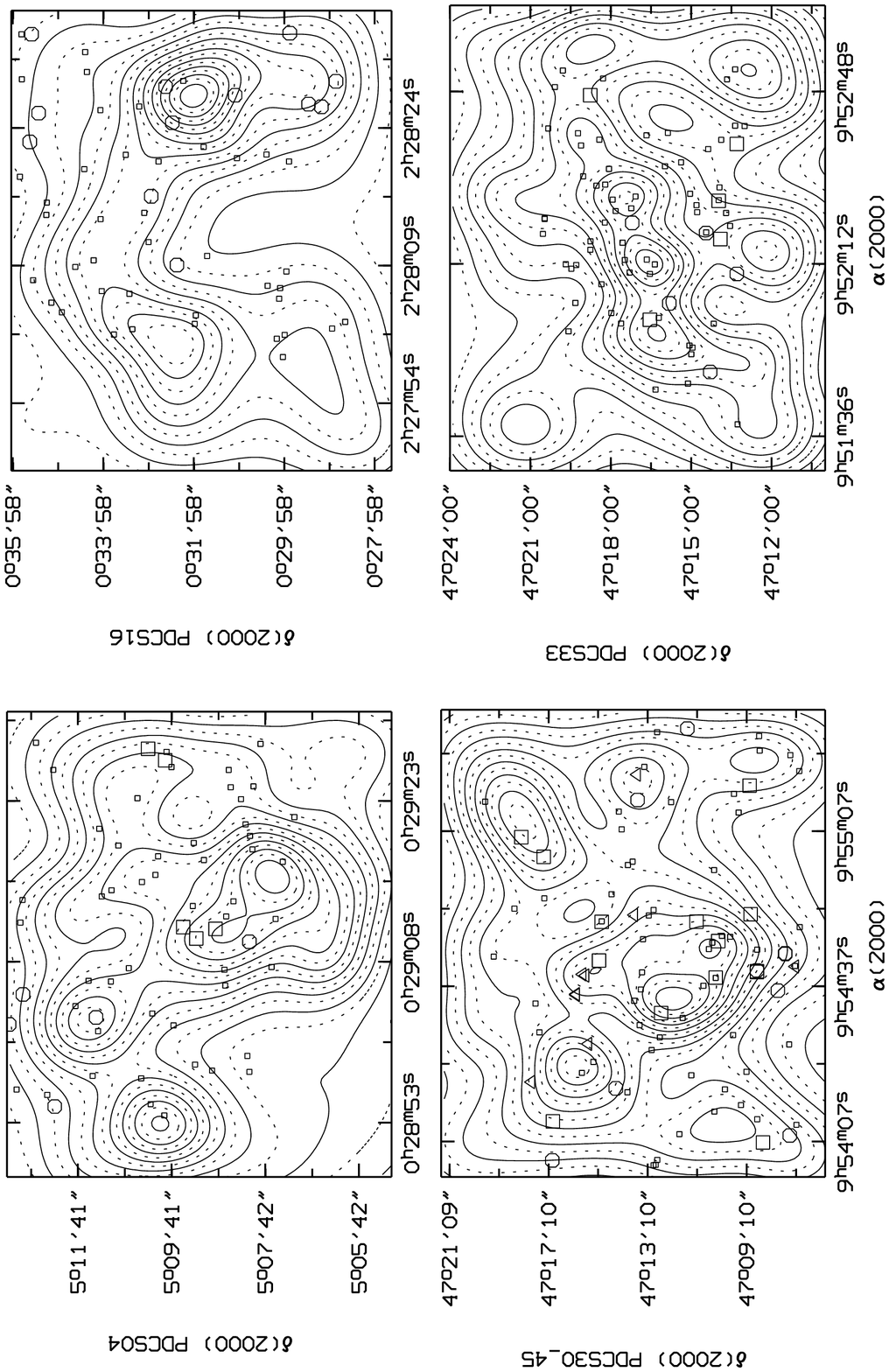]{upper left: galaxy isodensity contours for PDCS04. 
The position of the galaxies with a measured redshift are also plotted. The 
small squares are 
the field objects, the larger symbols (circles, squares, triangles) are the 
different structures defined in Tab. 4.; upper right: PDCS16 ; lower left:
PDCS30/45 ; lower right: PDCS33. We note that East is to the right and that 
some field galaxies appear to be very close to cluster galaxies, but we 
checked that they were not the same objects. \label{fig1}}

\figcaption[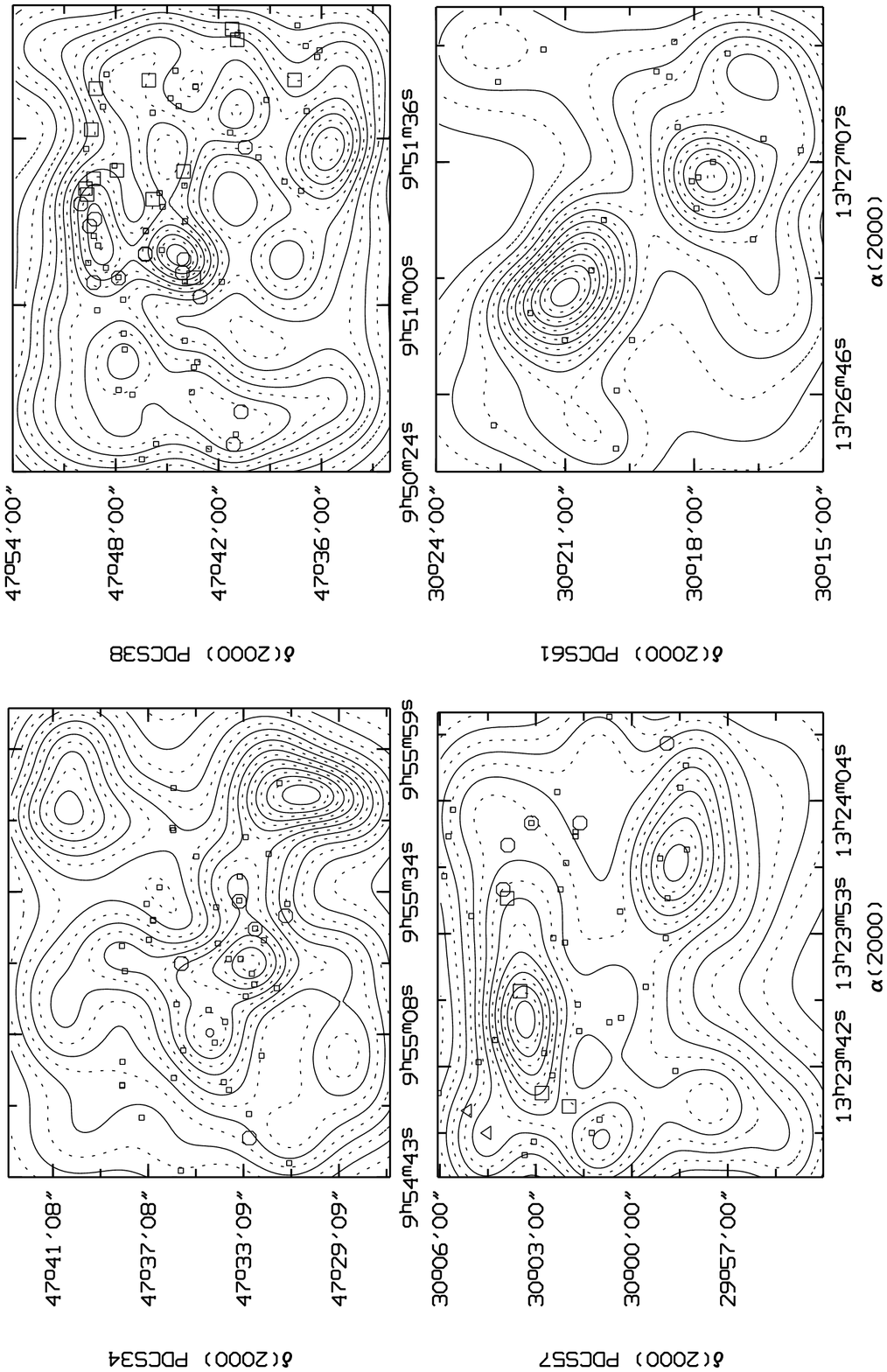]{upper left: PDCS34 ; upper right: PDCS38 ; lower 
left: PDCS57 ; lower right: PDCS61. \label{fig2}}

\figcaption[adami.f3.ps]{left: PDCS62 ; right: PDCS67 \label{fig3}}

\figcaption[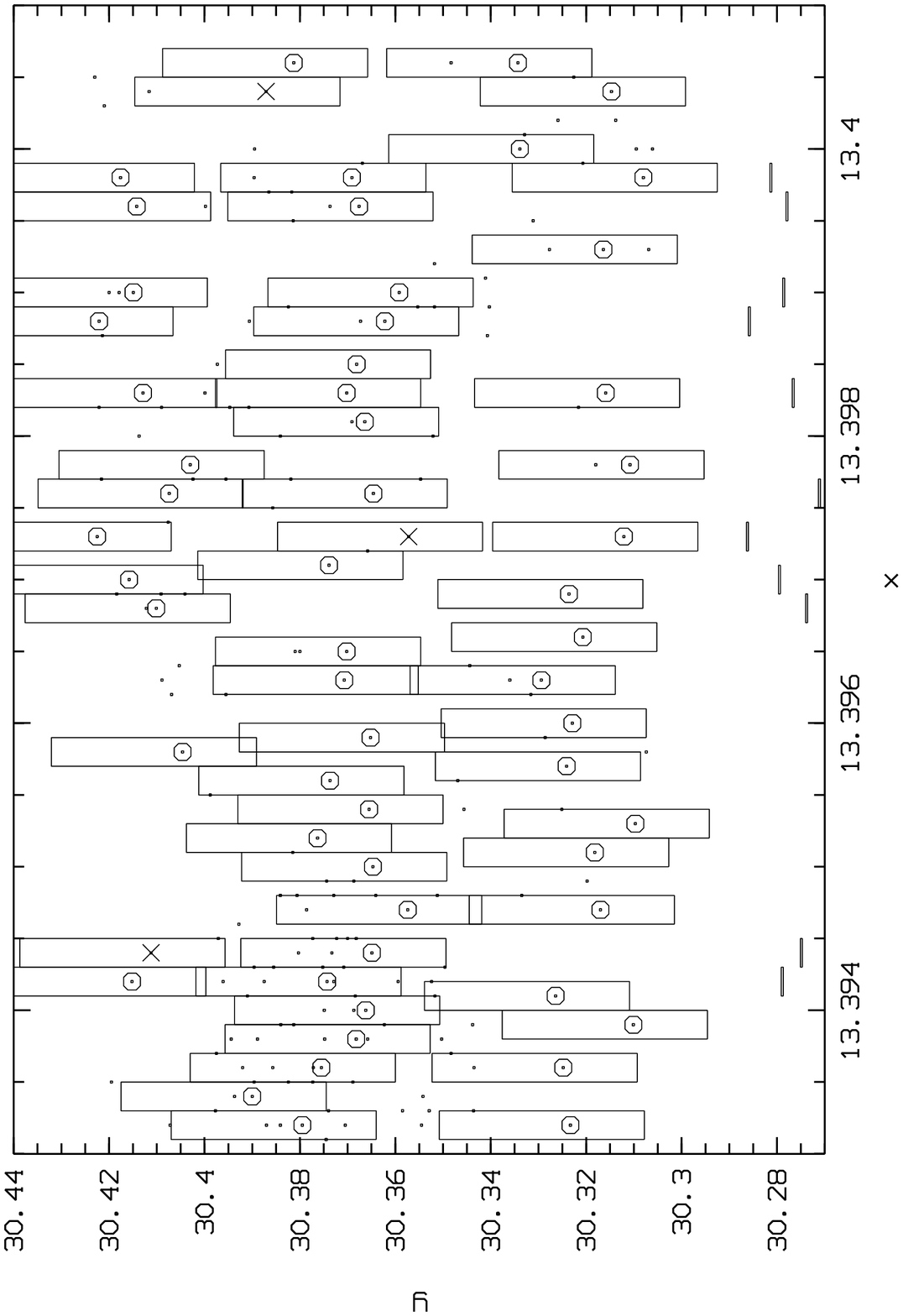]{Slit distribution for mask number 1 of PDCS62. The 
points 
are all the PDCS galaxies available in this field, the circled points and the
crosses are the spectroscopic targets. Circles denote primary and secondary 
targets. Crosses are the tertiary targets. We show the
region occupied by the spectrum and the region where the zero-order of the
spectrum lies on the CCD (thick small rectangles along the bottom of the
figure, visible only for the galaxies higher than 30.405 deg in declination). 
The x-axis is in hours and the y-axis is in degres (equinox 2000).
\label{fig4}}

\figcaption[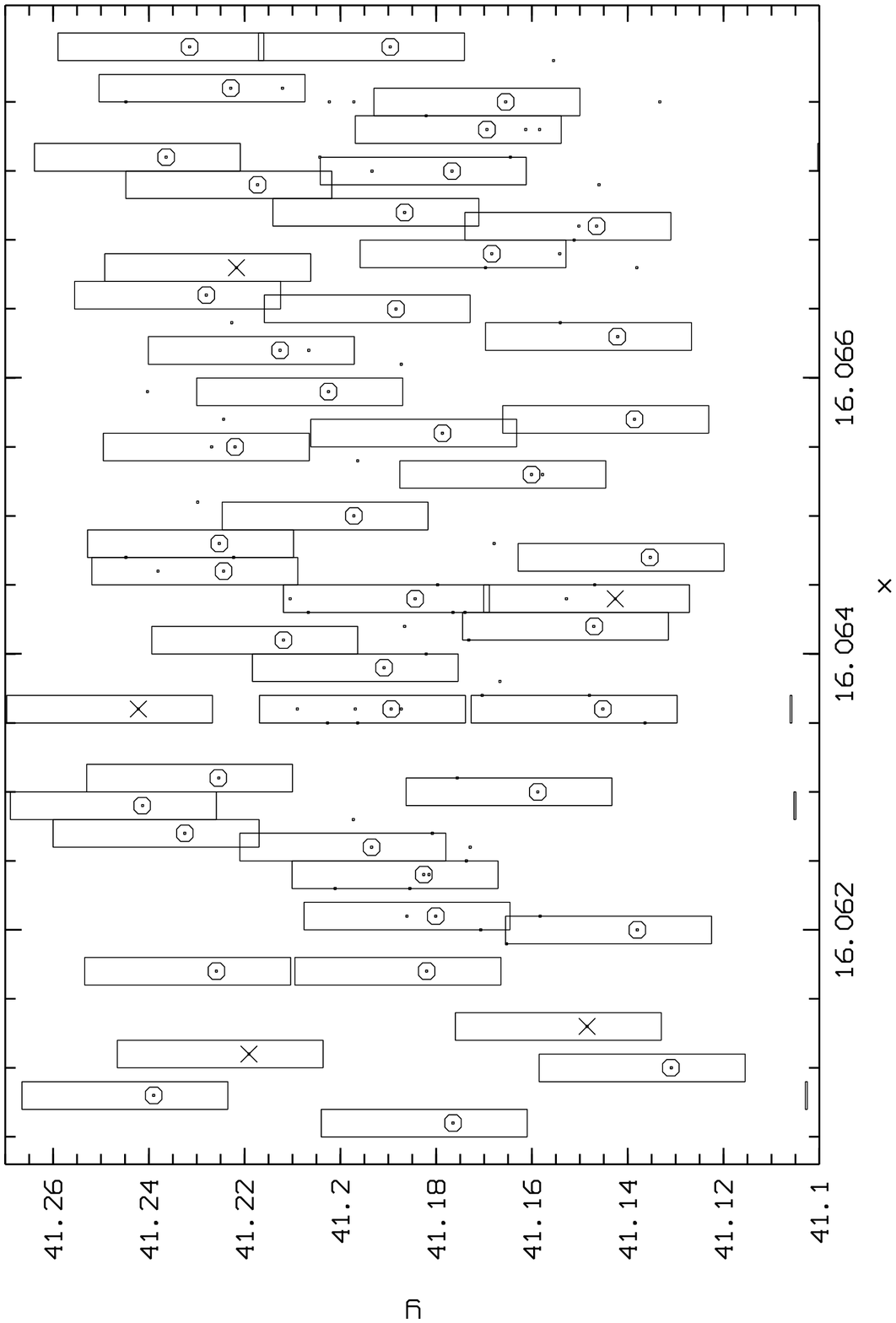]{Slit distribution for mask number 1 of PDCS67. The points 
are all
the PDCS galaxies available in this field, the circled points and the
crosses are the spectroscopic targets. Circles denote primary and secondary 
targets. Crosses are the tertiary targets. We show the region
occupied by the spectrum and the region where the zero-order of the
spectrum lies on the CCD (thick small rectangles along the bottom of the
figure, visible only for the galaxies higher than 41.215 deg in declination). 
The x-axis is in hours and the y-axis is in degres (equinox 2000). There are 
fewer slitlets then in Fig. 1 due to the lower target density.
\label{fig5}}

\figcaption[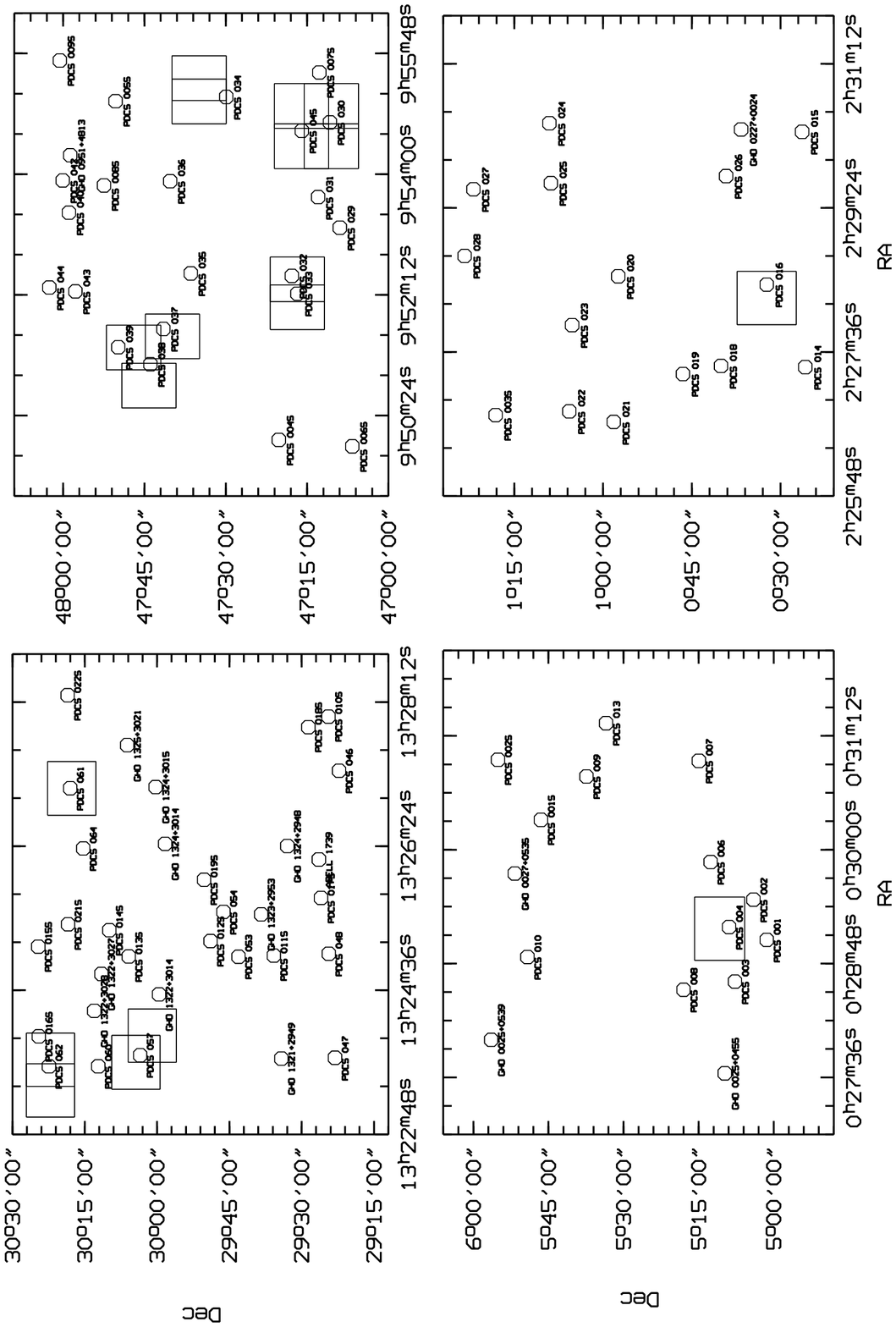]{The 9 PDCS lines of sight which are the most sampled by the 
COP 
survey. We note that the East is to the right. The x-axes are in hours and the 
y-axes in degres (equinox 2000). Each small box in the four fields is the area
sampled by one spectroscopic mask. The small labelled circles are the
position of the candidate clusters of galaxies as given in Postman et al.
(1996). \label{fig6}}

\figcaption[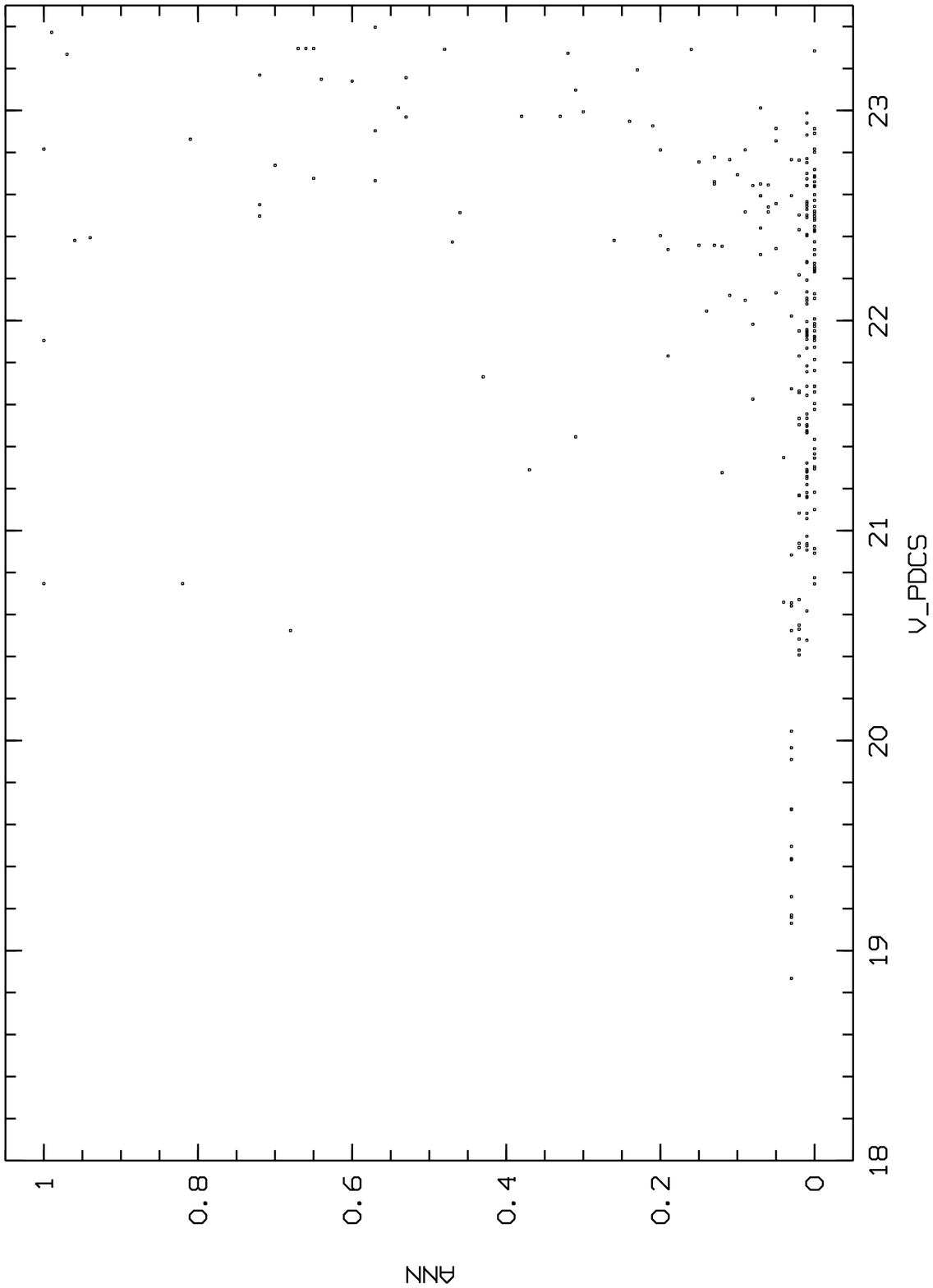]{Sextractor star-galaxy separation $only$ for the objects
classified as $galaxies$ by the PDCS for PDCS62 (all the points on the
figure are galaxies according to the PDCS: this gives raise to the
narrow distribution at the bright end). The y-axis is the ANN parameter
from sextractor. A value close to 0 means that the object is likely to be
a galaxy and a value close to 1 means that the object is likely to be a star
(according now to Sextractor). The x-axis is the $V_{PDCS}$ magnitude. We
see that the PDCS and Sextractor classifications only disagree partially at 
faint magnitudes.
\label{fig7}}

\figcaption[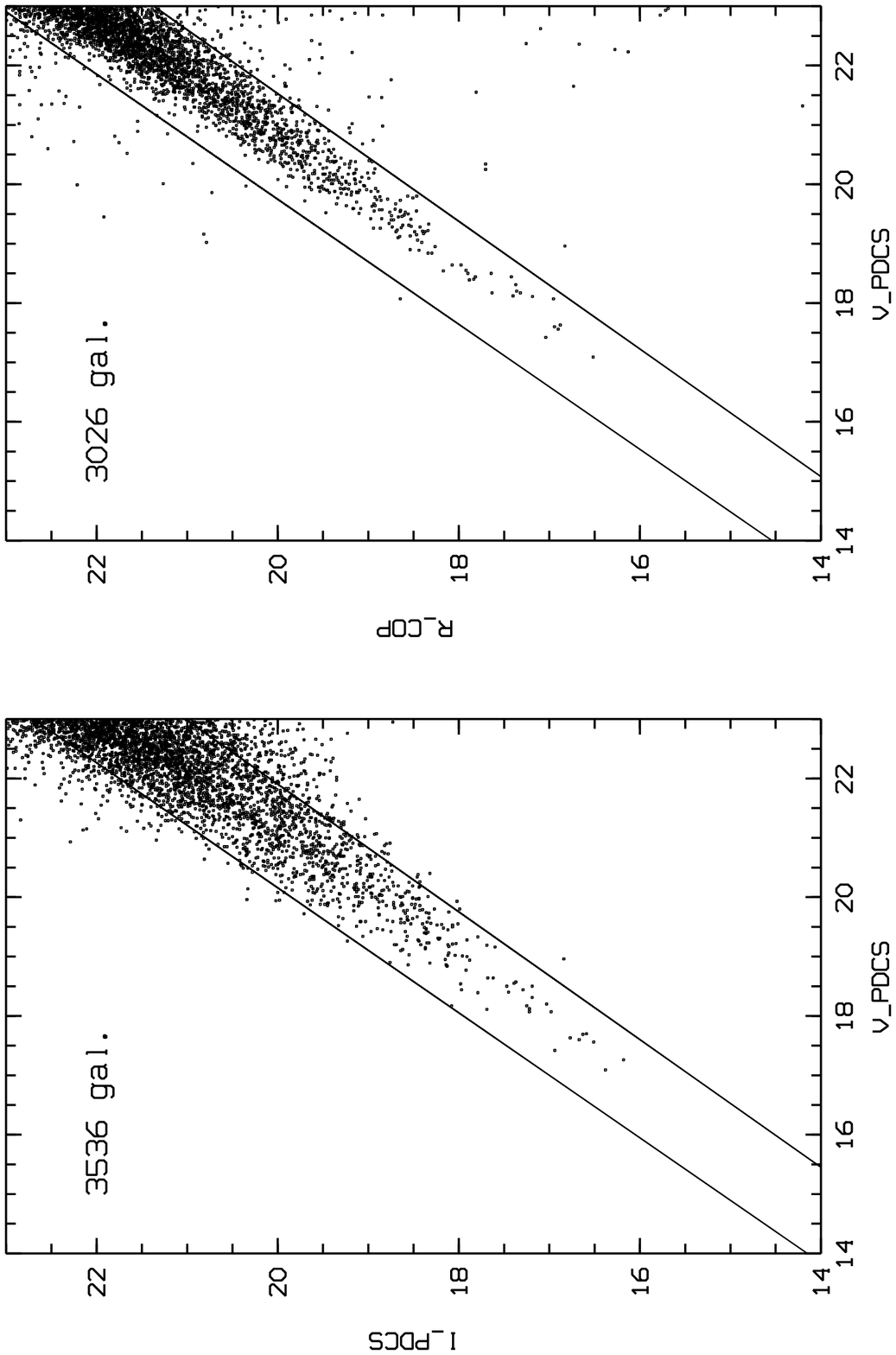]{Relations between the V, I and R magnitudes for 
the 10 lines of sight of this paper. The two x-axis are the $V_{PDCS}$ 
magnitude. The left yaxis is the $I_{PDCS}$ magnitude and the right
y-axis is the  $R_{COP}$ magnitude. The number of galaxies used in each
graph is indicated. The two straight lines show the 1-$\sigma$ envelope of the
relations. \label{fig8}}

\figcaption[adami.f9.ps]{The lower left spectrum represents a galaxy at 
z=0.4621$\pm$0.0002 with a cross-correlation coefficient of 13.26. The lower 
right spectrum is a galaxy where we have used emission lines to deduce the 
redshift (z=0.6579$\pm$0.0004), e.g. the [OII] line at
$\sim$6180\AA. The cross-correlation method detects also the H$\&$K lines 
around 6600\AA. The upper spectra (left and right) are two galaxies at 
z=0.4610$\pm$0.0003 (cross-correlation coefficient of 4.33) and at 
z=0.4587$\pm$0.0004 with both absorption line features (cross-correlation 
coefficient of 3.54) and emission line features ([OII]). The caption of each 
spectrum is the name of the line of sight (for example PDCS62) and the 
sequence of the spectrum in the observation process (for example 48) plus the
mask number (for example .1 for the first mask of PDCS62). \label{fig9}}

\figcaption[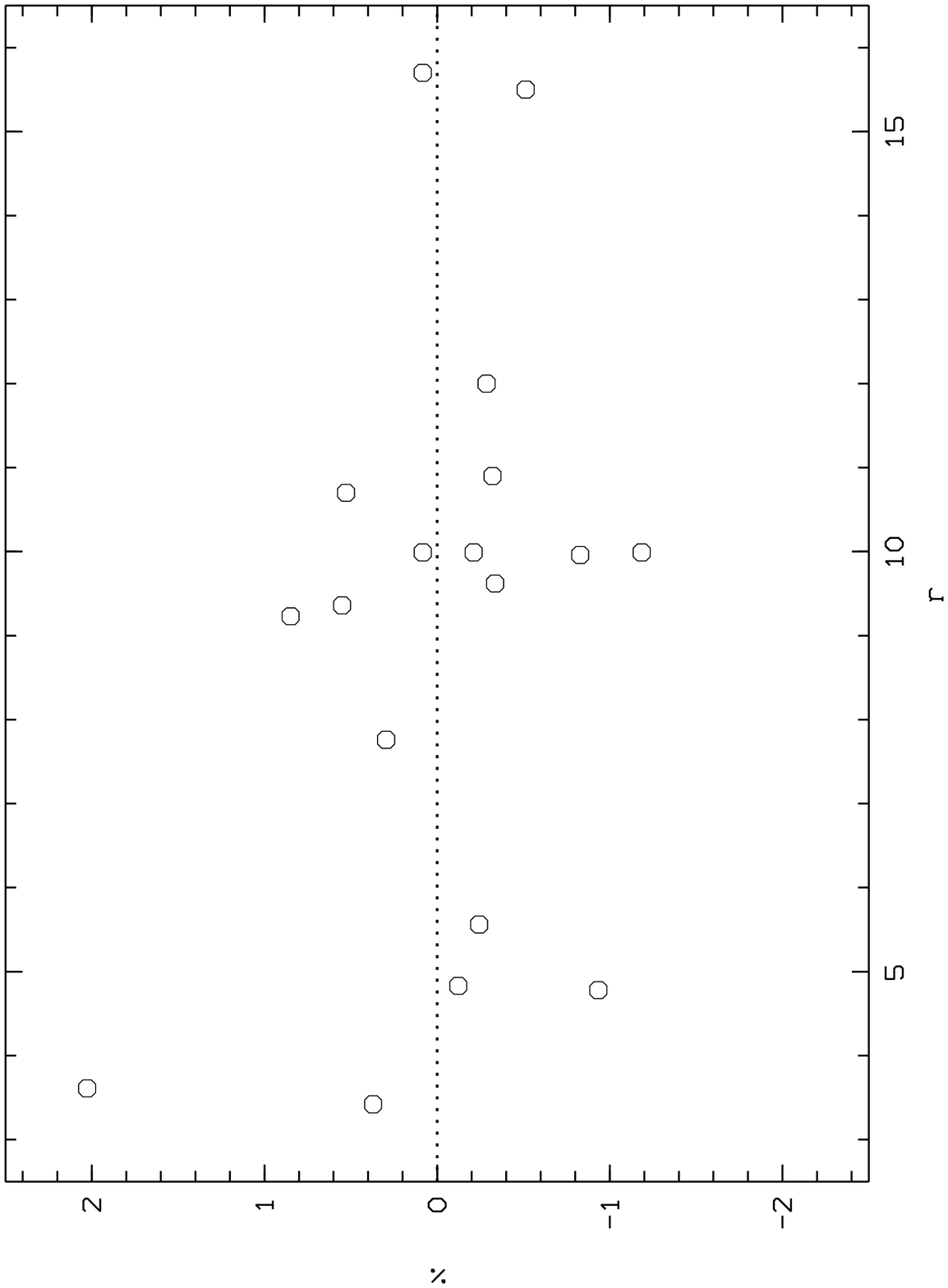]{Error percentage of the redshift estimation for 18 
galaxies observed twice. The zero level is symbolized with a 
thick dotted line. The x-axis is the Beers $\&$ Tonry cross- correlation 
coefficient computed with RVSAO and the y-axis is the percentage of 
difference between the two estimations of the redshift. \label{fig10}}

\figcaption[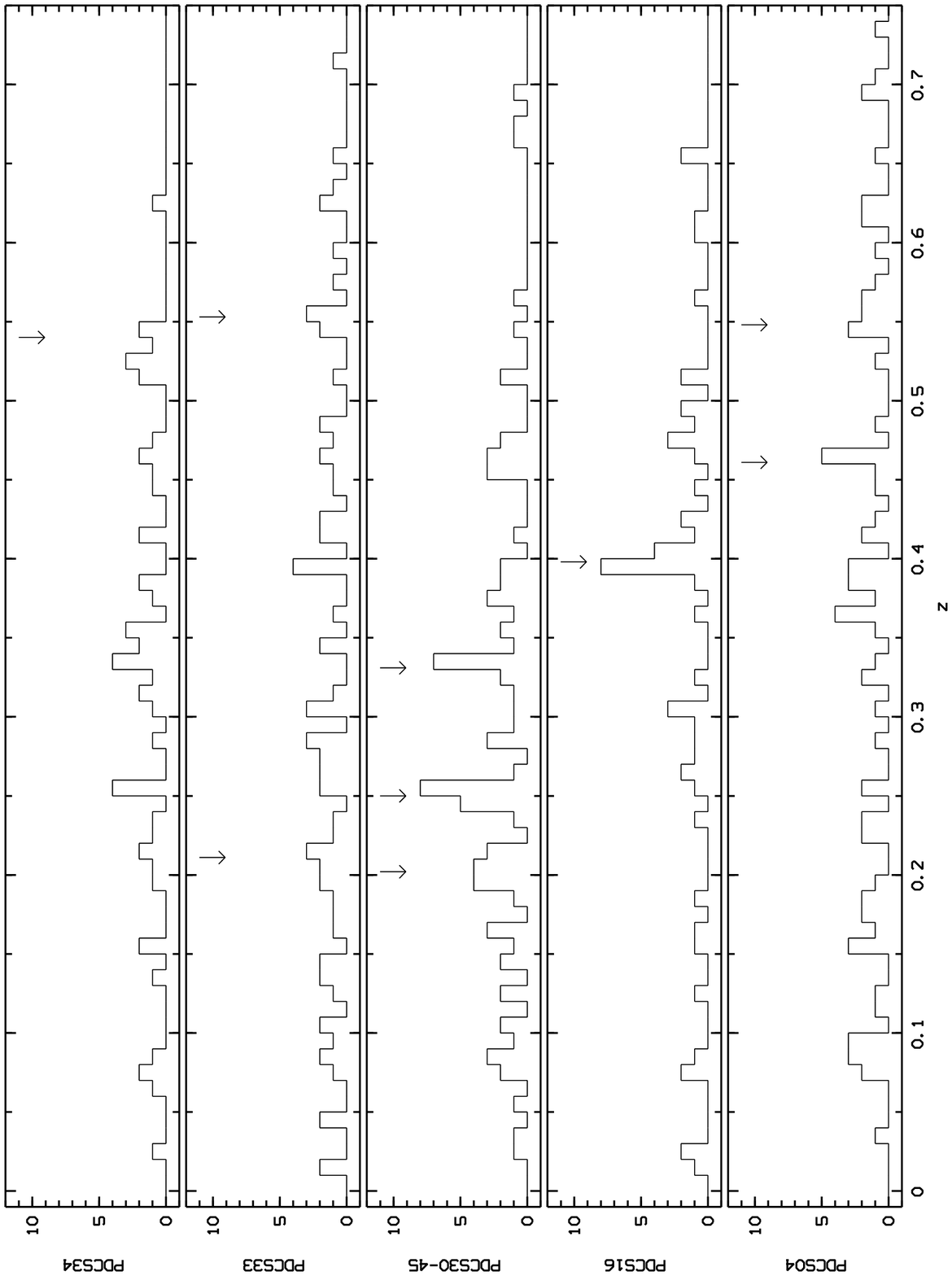]{Redshift distribution along the lines of sight of 
PDCS04, PDCS16,
PDCS30-45, PDCS33 and PDCS34. The x-axis is the redshift and the y-axis, the
number of redshifts. The significant structures (detected with the
gap method described in Section 5.1) are marked with an arrow. \label{fig11}}

\figcaption[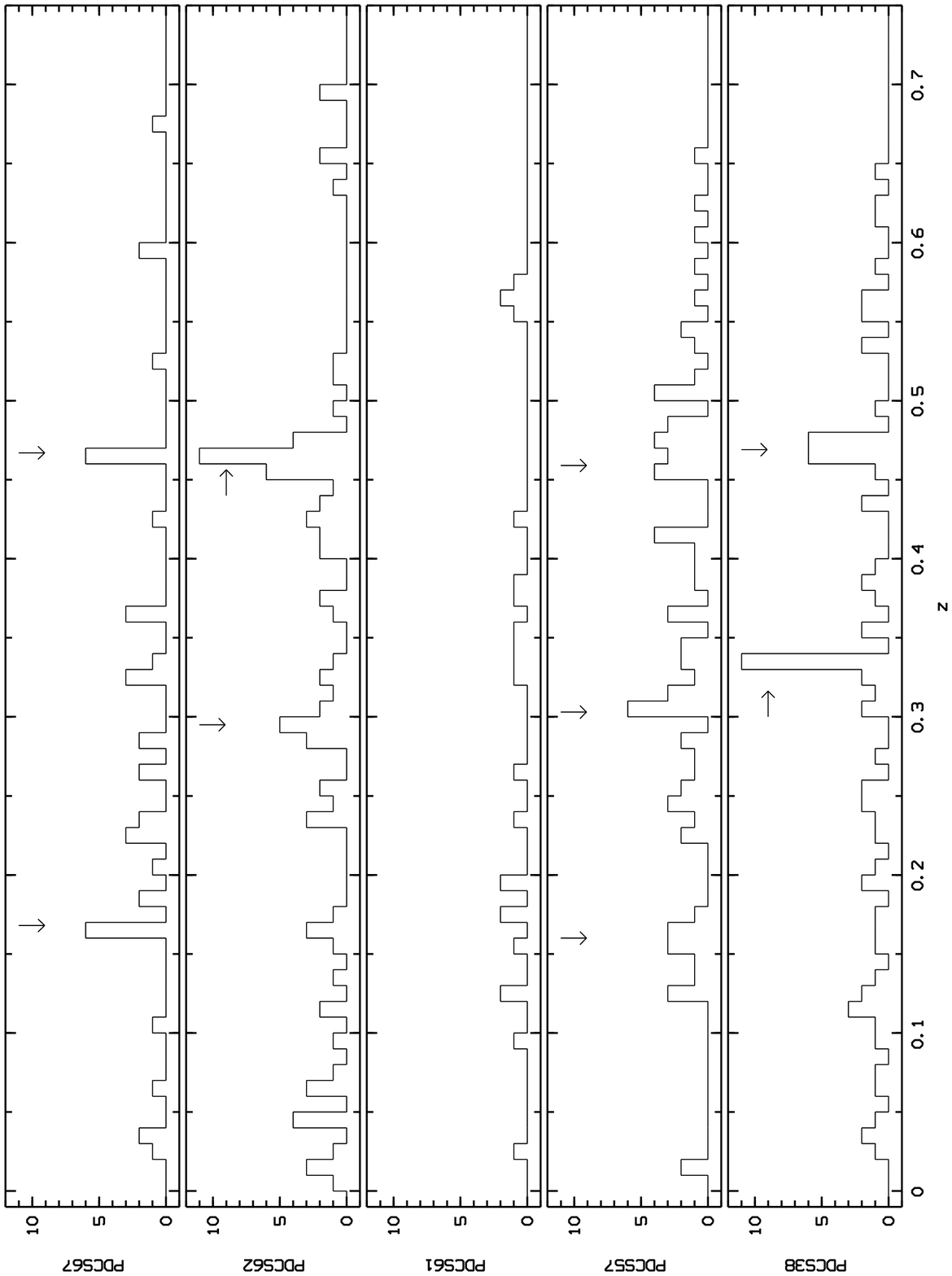]{Same caption as for Fig. 13, but for PDCS38, PDCS57, 
PDCS61, PDCS62 and PDCS67. \label{fig12}}

\figcaption[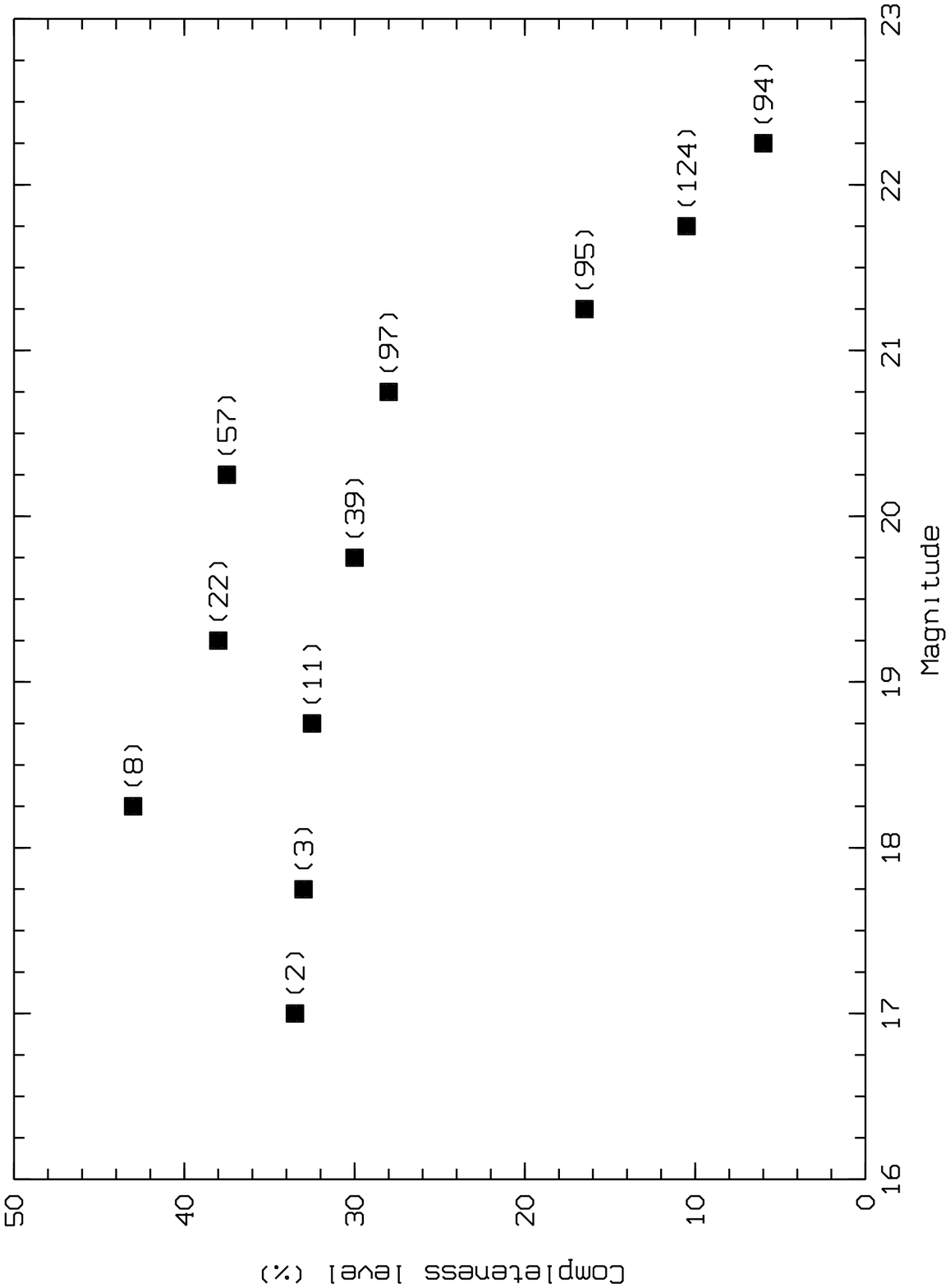]{Figure of the variation of the completeness level 
$C$ as a function of the $V_{PDCS}$ magnitude. $C$ is the ratio of the number
of galaxies with a measured redshift (given inside parentheses) and of the 
total number of galaxies in the fields.  We note that for the magnitudes 
fainter than 22.5, the completeness level is very low and not plotted on this 
figure. \label{fig13}}

\clearpage

\begin{deluxetable}{cccc}
\footnotesize
\tablecaption{Transmission level, spectral range and redshift range for
the 2 blocking filters. We note that filter $2503$ is limited on the blue 
side by the 
spectral range of STIS2: $\sim$4400\AA . The redshift range is computed to
include [OII], H$\&$K and the G band in the spectrum. \label{tbl-1}}
\tablewidth{0pt}
\tablehead{
\colhead{CFH number} & \colhead{Transmission}   & \colhead{wavelength
range}
 & \colhead{redshift range} 
}
\startdata
$2503$ & 83$\%$ & [(4089)4400;6835]\AA  & [0.20;0.56] \\ 
$4611$ & 93$\%$ & [5262;7042]\AA  & [0.43;0.61] \\
 \enddata
\end{deluxetable}

\clearpage

\begin{deluxetable}{cccccccccc}
\footnotesize
\tablecaption{Name of the PDCS cluster, number of redshifts along the line
of sight, number of masks, blocking filter, percentage of Emission Line
Galaxies along the line of sight, exposure time in minutes and 3-$\sigma$ 
X-ray emission detection. The last three columns are the V$_{PDCS}$ magnitude 
ranges where we selected the primary (1$^{st}$), secondary (2$^{nd}$) and tertiary targets (3$^{rd}$). We give the success rate 
(percentage of targets with a successfully measured redshift) for the
first magnitude bin, except for PDCS61 where the exposure time was 
drastically lower than expected. \label{tbl-2}}
\tablewidth{0pt}
\tablehead{
\colhead{Name} & \colhead{z}   & \colhead{mask} &
\colhead{filter} & \colhead{ELG} 
& \colhead{time} & \colhead{X-ray} 
& \colhead{1$^{st}$} & \colhead{2$^{nd}$} & \colhead{3$^{rd}$}
}
\startdata
PDCS04 & 74 & 2 & $4611$ & 15$\%$ & 120m & -  & [-;22.3] (65$\%$) & 
[22.3;22.8] & [22.8;23] \\ 
PDCS16 & 53 & 2 & $4611$ & 13$\%$ & 120m & -  & [-;22.3] (42$\%$) & 
[22.3;22.8] & [22.8;23] \\ 
PDCS30/45 & 88 & 4 & $2503$ & 23$\%$ & 90m & -  & [-;21.2] (55$\%$) & 
[21.2;21.7] & [21.7;23] \\ 
PDCS33/32 & 71 & 2 & $4611$ & 30$\%$ & 120m & yes  & [-;22.3] (71$\%$) & 
[22.3;22.8] & [22.8;23] \\ 
PDCS34 & 51 & 2 & $2503$ & 20$\%$ & 100m & -  & [-;21.5] (88$\%$) & 
[21.5;22] & [22;23] \\ 
PDCS38/39 & 75 & 3 & $2503$ & 29$\%$ & 80m & -  & [-;21.1] (62$\%$) & 
[21.1;21.6] & [21.6;23] \\ 
PDCS57 & 77 & 2 & $4611$ & 32$\%$ & 100m & -  & [-;22.1] (66$\%$) & 
[22.1;22.6] & [22.6;23] \\ 
PDCS61 & 22 & 1 & $2503$ & 25$\%$ & 20m & yes  & [-;21.1] (-) & 
[21.1;21.6] & [21.6;23] \\ 
PDCS62 & 84 & 2 & $4611$ & 25$\%$ & 100m & yes  & [-;22.3] (81$\%$) & 
[22.3;22.8] & [22.8;23] \\ 
PDCS67 & 41 & 1 & $4611$ & 39$\%$ & 180m & -  & [-;22.3] (60$\%$) & 
[22.3;22.8] & [22.8;23] \\ 
 \enddata
\end{deluxetable}

\clearpage

\begin{deluxetable}{cccccc}
\footnotesize
\tablecaption{Table of the images observed at CFH with number of
exposures, filter, exposure time, observation date and seeing.\label{tbl-3}}
\tablewidth{0pt}
\tablehead{
\colhead{Name} & \colhead{field}   & \colhead{filter} &
\colhead{time} & \colhead{Date} & \colhead{seeing}
}
\startdata
PDCS04 & 1 & $R_{COP}$ & 5mn & 08/98 & 1.0" \\ 
PDCS16 & 1 & $R_{COP}$ & 7+8mn & 08/98 & 1.0"+1.0" \\ 
PDCS30-45 & 1 & $R_{COP}$ & 5mn & 02/98 & 1.0" \\ 
PDCS30-45 & 2 & $R_{COP}$ & 5mn & 02/98 & 1.0" \\ 
PDCS30-45 & 3 & $R_{COP}$ & 5mn & 02/98 & 1.3" \\ 
PDCS30-45 & 3 & $2503$ & 5mn & 02/98 & 1.2" \\ 
PDCS30-45 & 4 & $R_{COP}$ & 5mn & 02/98 & 1.0" \\ 
PDCS33-32 & 1 & $R_{COP}$ & 5mn & 02/98 & 1.8" \\ 
PDCS33-32 & 1 & $4611$ & 5mn & 02/98 & 0.9" \\ 
PDCS33-32 & 2 & $4611$ & 5mn & 02/98 & 0.8" \\ 
PDCS34 & 1 & $R_{COP}$ & 5mn & 02/98 & 1.4" \\ 
PDCS34 & 1 & $2503$ & 5mn & 02/98 & 1.1" \\ 
PDCS34 & 2 & $2503$ & 5mn & 02/98 & 1.1" \\ 
PDCS37-38-39 & 1 & $R_{COP}$ & 8mn & 02/98 & 0.8" \\
PDCS37-38-39 & 2 & $R_{COP}$ & 8mn & 02/98 & 0.9" \\ 
PDCS37-38-39 & 3 & $R_{COP}$ & 5mn & 02/98 & 1.1" \\ 
PDCS57 & 1 & $R_{COP}$ & 5mn & 02/98 & 0.8" \\ 
PDCS57 & 1 & $4611$ & 5mn & 02/98 & 1.1" \\ 
PDCS57 & 2 & $4611$ & 5mn & 02/98 & 1.0" \\ 
PDCS61 & 1 & $2503$ & 5mn & 02/98 & 1.0" \\ 
PDCS62 & 1 & $R_{COP}$ & 5mn & 02/98 & 0.9" \\ 
PDCS62 & 1 & $V_{COP}$ & 5mn & 02/98 & 1.0" \\ 
PDCS62 & 1 & $4611$ & 5mn & 02/98 & 1.0" \\ 
PDCS62 & 2 & $V_{COP}$ & 5mn & 02/98 & 1.2" \\ 
PDCS62 & 2 & $4611$ & 5mn & 02/98 & 1.0" \\ 
PDCS67 & 1 & $R_{COP}$ & 5mn & 08/98 & 2.1" \\
 \enddata
\end{deluxetable}

\clearpage

\begin{deluxetable}{ccccc}
\footnotesize
\tablecaption{Table of the detected structures along the 10 PDCS lines
of sight. The columns are the name of the line of sight ($los$), the
structure number, the mean redshift of this structure, the total number of 
redshifts in each structure and the number of emission lines galaxies in
these
structures are given. We note that the structures of this table are
sampled with more than 5 galaxies. \label{tbl-4}}
\tablewidth{0pt}
\tablehead{
\colhead{$los$} & \colhead{struct.}   & \colhead{mean z} &
\colhead{number of z} & \colhead{em. lines galaxies}
}
\startdata
PDCS04 & 1 & 0.4608 & 5 & 0 \\ 
PDCS04 & 2 & 0.5486 & 5 & 0 \\ 
PDCS16 & 1 & 0.3984 & 12 & 0 \\ 
PDCS30-45 & 1 & 0.2017 & 9 & 1 \\ 
PDCS30-45 & 2 & 0.2500 & 14 & 4 \\ 
PDCS30-45 & 3 & 0.3312 & 9 & 1 \\ 
PDCS33 & 1 & 0.2106 & 5 & 2 \\ 
PDCS33 & 2 & 0.5530 & 5 & 3 \\ 
PDCS34 & 1 & 0.5541 & 5 & 0 \\ 
PDCS38 & 1 & 0.3316 & 13 & 4 \\
PDCS38 & 2 & 0.4691 & 13 & 4 \\ 
PDCS57 & 1 & 0.1605 & 6 & 5 \\ 
PDCS57 & 2 & 0.3033 & 6 & 2 \\ 
PDCS57 & 3 & 0.4586 & 7 & 2 \\ 
PDCS62 & 1 & 0.2948 & 5 & 0 \\ 
PDCS62 & 2 & 0.4619 & 19 & 3 \\ 
PDCS67 & 1 & 0.1686 & 6 & 0 \\
PDCS67 & 2 & 0.4675 & 6 & 4 \\
 \enddata
\end{deluxetable}

\clearpage

\begin{deluxetable}{ccccccc}
\footnotesize
\tablecaption{Table of the spatial representativity levels of the sample of 
galaxies with a measured redshift compared to the sample of all the PDCS 
galaxies (as a function of the limiting magnitude). We have used a 
bidimensional Kolmogorov-Smirnov test as a function of the V limiting 
magnitude. A value close to 100 means that the two distributions (sample with 
redshift $versus$ total sample) have a very similar spatial distribution. A 
value lower than 90 means that the 2 distributions are spatially differently
distributed at the level of 10$\%$.  We have tested each of the masks 
of the PDCS30/45 line of sight to see if the low representativity level of 
this line of sight was due to one peculiar mask. We give also for each 
representativity level the number of galaxies with a measured redshift 
(only when we used more than 10 galaxies). \label{tbl-5}}
\tablewidth{0pt}
\tablehead{
\colhead{} &  \colhead{$<$20.} &
\colhead{$<$20.5} & \colhead{$<$21.} & \colhead{$<$21.5} & \colhead{$<$22.}
 & \colhead{$<$22.5}
}
\startdata
PDCS04 &  - & 99.9 (10) & 99.9 (22) & 99.9 (36) & 99.9 (49) & 99.9 (63) \\ 
PDCS16 &  - & - & 99.9 (11) & 99.9 (21) & 99.9 (34) & 65.9 (42) \\ 
PDCS30-45 all &  61.0 (17) & 58.0 (32) & 48.0 (49) & 24.9 (57) & 30.4 (72) & 
14.5 (81) \\ 
PDCS30-45 1 &  - & - & 57.7 (12) & 11.5 (13) & 2.8 (17) & 2.2 (21) \\ 
PDCS30-45 2 &  - & 76.5 (14) & 46.4 (19) & 22.7 (24) & 37.6 (27) & 36.6 (28) \\ 
PDCS30-45 3 &  - & - & - & - & 77.0 (12) & 86.7 (13) \\ 
PDCS30-45 4 &  - & - & 71.8 (12) & 30.8 (12) & 16.9 (16) & 3.6 (19) \\
PDCS33 & - &  99.9 (13) & 99.9 (19) & 99.9 (26) & 99.9 (36) & 68.3 (48) \\
\enddata
\end{deluxetable}

\clearpage

\begin{deluxetable}{cccccccccc}
\footnotesize
\tablecaption{Data for the line of sight PDCS04.\label{tbl-6}}
\tablewidth{0pt}
\tablehead{
\colhead{Num.} & \colhead{group}   & \colhead{z}
& \colhead{err-z}   & \colhead{Alpha} & \colhead{Delta}
& \colhead{V} & \colhead{I} & \colhead{R} & \colhead{type}
}
\startdata
58.1 &  0 &  0.03822 &  0.00016 &   26047.6 &   18830.9 &  99.99
 &  99.99 &  17.17 &   \\
46.1 &  0 &  0.07219 &  0.00001 &   26524.6 &   18754.9 &  20.55
 &  19.70 &  19.95 & ea\\
32.1 &  0 &  0.07864 &  0.00045 &   26447.5 &   18625.1 &  21.32
 &  21.98 &  20.40 &   \\
36.2 &  0 &  0.08469 &  0.00025 &   26193.2 &   18687.7 &  22.37
 &  21.09 &  21.61 &   \\
 5.1 &  0 &  0.08489 &  0.00018 &   26464.0 &   18476.7 &  22.95
 &  22.28 &  22.09 &   \\
45.2 &  0 &  0.08958 &  0.00027 &   26287.0 &   18838.3 &  18.90
 &  18.76 &  18.54 &   \\
38.1 &  0 &  0.09029 &  0.00042 &   26123.0 &   18675.5 &  22.35
 &  20.94 &  20.92 &   \\
20.2 &  0 &  0.09804 &  0.00046 &   26511.9 &   18587.9 &  20.99
 &  20.54 &  20.23 &   \\
\enddata
\tablenotetext{a}{ 1$^{st}$ column: observed galaxy sequence (58.1 means 
58$^{th}$ galaxy observed in the mask 1). 2$^{nd}$ column: field/group 
classification, 0 means field galaxy, 1 means first group ..etc... 3$^{rd}$ 
column: redshift (the last digit is given only for information, but is not
significant according to the error). 4$^{th}$ column: redshift error. 5$^{th}$ 
column: alpha coordinates from the CFH observations in arcsec (equinox 2000). 
6$^{th}$ column: delta coordinates from the CFH observations in arcsec 
(equinox 2000). 7$^{th}$ column: PDCS V magnitude. 8$^{th}$ column: PDCS I 
magnitude. 9$^{th}$ column: CFH R magnitude. 10$^{th}$ column: spectral 
features: "em" means redshift from only emission lines, "ea" means redshift 
from emission lines but consistent with absorption line features, " " means 
redshift only from absorption line features. When a magnitude was not available
or not successfully measured, we put the number 99.99.}

\end{deluxetable}

\clearpage

\begin{deluxetable}{cccccccccc}
\footnotesize
\tablecaption{Data for the line of sight PDCS16.\label{tbl-7}}
\tablewidth{0pt}
\tablehead{
\colhead{Num.} & \colhead{group}   & \colhead{z}
& \colhead{err-z}   & \colhead{Alpha} & \colhead{Delta}
& \colhead{V} & \colhead{I} & \colhead{R} & \colhead{type}
}
\startdata
2.35 &  0 &  0.01236 &  0.00024 &  133519.9 &    2008.3 &  18.07
 &  17.22 &  18.65 &   \\
2.30 &  0 &  0.02487 &  0.00019 &  133484.2 &    2148.5 &  21.07
 &  19.59 &  20.22 &   \\
2.51 &  0 &  0.02856 &  0.00021 &  133654.9 &    2059.6 &  21.35
 &  20.30 &  20.32 &   \\
1.20 &  0 &  0.07260 &  0.00016 &  133377.8 &    1978.0 &  21.71
 &  21.30 &  21.45 &   \\
1.52 &  0 &  0.07968 &  0.00036 &  133640.3 &    1932.0 &  21.40
 &  19.52 &  20.09 &   \\
1.45 &  0 &  0.08275 &  0.00022 &  133592.8 &    2041.8 &  21.32
 &  20.33 &  20.71 &   \\
1.15 &  0 &  0.12853 &  0.00034 &  133302.8 &    1803.6 &  22.15
 &  21.74 &  22.25 & em \\
2.49 &  0 &  0.15218 &  0.00026 &  133643.0 &    2145.9 &  21.53
 &  21.16 &  21.27 & em \\
\enddata
\end{deluxetable}

\clearpage

\begin{deluxetable}{cccccccccc}
\footnotesize
\tablecaption{Data for the line of sight PDCS30/45.\label{tbl-8}}
\tablewidth{0pt}
\tablehead{
\colhead{Num.} & \colhead{group}   & \colhead{z}
& \colhead{err-z}   & \colhead{Alpha} & \colhead{Delta}
& \colhead{V} & \colhead{I} & \colhead{R} & \colhead{type}
}
\startdata
1.43 &  0 &  0.02042 &  0.00041 &  535818.8 &  169644.2 &  21.63
 &  20.52 &  21.00 &    \\
2.45 &  0 &  0.03008 &  0.00021 &  535322.0 &  169621.6 &  20.76
 &  20.32 &  20.39 &    \\
32.4 &  0 &  0.05074 &  0.00006 &  535101.1 &  170261.3 &  20.45
 &  20.13 &  19.85 & em \\
2.15 &  0 &  0.07501 &  0.00060 &  534807.9 &  169763.8 &  19.59
 &  19.09 &  19.17 &    \\
1.11 &  0 &  0.07601 &  0.00022 &  535276.6 &  169833.2 &  20.74
 &  19.69 &  20.03 &    \\
3.18 &  0 &  0.08746 &  0.00034 &  535375.4 &  170312.4 &  21.63
 &  19.88 &  20.38 &    \\
2.37 &  0 &  0.08779 &  0.00054 &  535191.3 &  170003.5 &  99.99
 &  99.99 &  17.05 &    \\
2.28 &  0 &  0.08841 &  0.00152 &  535009.9 &  169782.8 &  20.54
 &  19.02 &  19.68 &    \\
\enddata
\end{deluxetable}

\clearpage

\begin{deluxetable}{cccccccccc}
\footnotesize
\tablecaption{Data for the line of sight PDCS33.\label{tbl-9}}
\tablewidth{0pt}
\tablehead{
\colhead{Num.} & \colhead{group}   & \colhead{z}
& \colhead{err-z}   & \colhead{Alpha} & \colhead{Delta}
& \colhead{V} & \colhead{I} & \colhead{R} & \colhead{type}
}
\startdata
1.20 &  0 &  0.01166 &  0.00018 &  532697.0 &  170099.6 &  21.64
 &  21.05 &  21.07 &    \\
1.5  &  0 &  0.01961 &  0.00014 &  532477.3 &  169996.3 &  21.64
 &  19.91 &  20.21 &    \\
2.35 &  0 &  0.04537 &  0.00039 &  533225.2 &  170312.0 &  20.17
 &  18.79 &  19.22 &    \\
2.13 &  0 &  0.04622 &  0.00042 &  532964.9 &  170368.2 &  22.86
 &  21.52 &  21.95 &    \\
2.50 &  0 &  0.07506 &  0.00035 &  533369.3 &  170305.6 &  21.03
 &  19.25 &  99.99 &    \\
1.40 &  0 &  0.08026 &  0.00066 &  532953.0 &  170237.9 &  20.01
 &  19.14 &  19.39 &    \\
2.67 &  0 &  0.08865 &  0.00022 &  533558.3 &  170019.7 &  22.24
 &  20.88 &  99.99 &    \\
1.26 &  0 &  0.09783 &  0.00023 &  532768.9 &  170374.7 &  22.71
 &  22.10 &  22.41 &    \\
\enddata
\end{deluxetable}

\clearpage

\begin{deluxetable}{cccccccccc}
\footnotesize
\tablecaption{Data for the line of sight PDCS34.\label{tbl-10}}
\tablewidth{0pt}
\tablehead{
\colhead{Num.} & \colhead{group}   & \colhead{z}
& \colhead{err-z}   & \colhead{Alpha} & \colhead{Delta}
& \colhead{V} & \colhead{I} & \colhead{R} & \colhead{type}
}
\startdata
1.29 &  0 &  0.02708 &  0.00066 &  535659.5 &  171235.1 &  22.48
 &  21.33 &  21.51 &    \\
1.7  &  0 &  0.06390 &  0.00082 &  535285.3 &  171073.8 &  19.85
 &  19.11 &  19.30 &    \\
1.6  &  0 &  0.07104 &  0.00018 &  535265.3 &  171346.3 &  22.49
 &  20.75 &  21.62 &    \\
1.47 &  0 &  0.07899 &  0.00037 &  535971.6 &  171432.4 &  21.64
 &  20.71 &  21.03 &    \\
1.16 &  0 &  0.08948 &  0.00050 &  535415.4 &  171175.0 &  20.41
 &  19.83 &  19.84 &    \\
2.38 &  0 &  0.13132 &  0.00019 &  536015.9 &  171400.0 &  18.94
 &  17.88 &  99.99 &    \\
2.56 &  0 &  0.15736 &  0.00027 &  536173.2 &  171367.6 &  21.12
 &  20.06 &  99.99 &    \\
2.40 &  0 &  0.15866 &  0.00037 &  536044.5 &  171198.0 &  17.26
 &  16.18 &  99.99 &    \\
\enddata
\end{deluxetable}

\clearpage

\begin{deluxetable}{cccccccccc}
\footnotesize
\tablecaption{Data for the line of sight PDCS38.\label{tbl-11}}
\tablewidth{0pt}
\tablehead{
\colhead{Num.} & \colhead{group}   & \colhead{z}
& \colhead{err-z}   & \colhead{Alpha} & \colhead{Delta}
& \colhead{V} & \colhead{I} & \colhead{R} & \colhead{type}
}
\startdata
1.29 &  0 &  0.02885 &  0.00043 &  531785.0 &  171838.4 &  21.00
 &  19.54 &  20.02 & ea \\
3.20 &  0 &  0.03049 &  0.00027 &  532379.0 &  171580.7 &  22.06
 &  21.39 &  21.35 &    \\
3.25 &  0 &  0.03242 &  0.00032 &  532459.4 &  171679.7 &  22.62
 &  19.91 &  21.05 &    \\
1.47 &  0 &  0.04658 &  0.00029 &  531975.6 &  171708.8 &  99.99
 &  99.99 &  99.99 &    \\
3.30 &  0 &  0.06280 &  0.00026 &  532529.1 &  171411.8 &  21.87
 &  20.08 &  20.58 &    \\
2.15 &  0 &  0.07395 &  0.00021 &  532124.6 &  172156.7 &  17.63
 &  16.77 &  16.88 &    \\
1.6  &  0 &  0.09526 &  0.00020 &  531400.5 &  171988.2 &  22.51
 &  22.09 &  21.71 &    \\
1.27 &  0 &  0.10543 &  0.00019 &  531756.4 &  172048.3 &  21.21
 &  19.92 &  20.23 &    \\
\enddata
\end{deluxetable}

\clearpage

\begin{deluxetable}{cccccccccc}
\footnotesize
\tablecaption{Data for the line of sight PDCS57.\label{tbl-12}}
\tablewidth{0pt}
\tablehead{
\colhead{Num.} & \colhead{group}   & \colhead{z}
& \colhead{err-z}   & \colhead{Alpha} & \colhead{Delta}
& \colhead{V} & \colhead{I} & \colhead{R} & \colhead{type}
}
\startdata
1.17 &  0 &  0.01421 &  0.00018 &  723173.4 &  108354.6 &  99.99
 &  99.99 &  18.93 &    \\
2.30 &  0 &  0.01751 &  0.00068 &  723594.6 &  107896.7 &  19.71
 &  18.88 &  99.99 &    \\
1.64 &  0 &  0.12047 &  0.00001 &  723578.4 &  108123.1 &  21.55
 &  21.50 &  21.29 & em \\
2.42 &  0 &  0.12386 &  0.00041 &  723697.2 &  107898.5 &  21.39
 &  20.30 &  99.99 &    \\
2.3  &  0 &  0.12693 &  0.00025 &  723346.2 &  108164.2 &  21.48
 &  20.08 &  20.55 &    \\
2.51 &  0 &  0.13010 &  0.00010 &  723799.8 &  107872.9 &  22.31
 &  21.73 &  99.99 & em \\
2.24 &  0 &  0.14819 &  0.00041 &  723546.0 &  108133.6 &  21.77
 &  21.57 &  21.58 & em \\
2.35 &  1 &  0.15820 &  0.00050 &  723627.0 &  108188.3 &  21.85
 &  21.11 &  21.06 & em \\
\enddata
\end{deluxetable}

\clearpage

\begin{deluxetable}{cccccccccc}
\footnotesize
\tablecaption{Data for the line of sight PDCS61.\label{tbl-13}}
\tablewidth{0pt}
\tablehead{
\colhead{Num.} & \colhead{group}   & \colhead{z}
& \colhead{err-z}   & \colhead{Alpha} & \colhead{Delta}
& \colhead{V} & \colhead{I} & \colhead{R} & \colhead{type}
}
\startdata
1.40 &  0 &  0.02084 &  0.00053 &  726519.6 &  109353.6 &  99.99
 &  99.99 &  22.49 &    \\
1.6  &  0 &  0.09630 &  0.00021 &  726040.8 &  109359.7 &  22.35
 &  22.02 &  22.52 &    \\
1.49 &  0 &  0.12389 &  0.00029 &  726559.2 &  109033.2 &  19.14
 &  18.09 &  20.38 &    \\
1.29 &  0 &  0.12404 &  0.00045 &  726300.0 &  108997.9 &  19.54
 &  19.34 &  19.58 & em \\
1.37 &  0 &  0.15591 &  0.00057 &  726424.2 &  108931.3 &  20.85
 &  19.22 &  21.47 &    \\
1.34 &  0 &  0.17021 &  0.00015 &  726386.4 &  109074.2 &  19.42
 &  17.90 &  19.06 &    \\
1.51 &  0 &  0.17026 &  0.00037 &  726575.4 &  109107.7 &  20.14
 &  19.46 &  20.06 & em \\
1.50 &  0 &  0.19231 &  0.00025 &  726564.6 &  109290.2 &  22.20
 &  22.04 &  21.72 & em \\
\enddata
\tablenotetext{a}{The last galaxy of this table (1.42) exhibits a single
broad band emission line, probably from a quasar. The
cross-correlation technique provides, however, with the absorption line features
a redshift equal to 0.57801$\pm$0.00049.}

\end{deluxetable}

\clearpage

\begin{deluxetable}{cccccccccc}
\footnotesize
\tablecaption{Data for the line of sight PDCS62.\label{tbl-14}}
\tablewidth{0pt}
\tablehead{
\colhead{Num.} & \colhead{group}   & \colhead{z}
& \colhead{err-z}   & \colhead{Alpha} & \colhead{Delta}
& \colhead{V} & \colhead{I} & \colhead{R} & \colhead{type}
}
\startdata
63.2 &  0 &  0.00952 &  0.00091 &  723176.0 &  109521.6 &  17.24
 &  99.99 &  16.72 &    \\
37.2 &  0 &  0.01632 &  0.00028 &  723308.4 &  109370.2 &  22.32
 &  21.25 &  21.62 &    \\
63.1 &  0 &  0.01801 &  0.00024 &  723513.6 &  109489.3 &  22.93
 &  21.46 &  22.20 &    \\
 7.1 &  0 &  0.01806 &  0.00019 &  723254.4 &  109116.4 &  22.16
 &  21.44 &  21.49 &    \\
12.2 &  0 &  0.02595 &  0.00033 &  723034.0 &  109173.9 &  22.45
 &  99.99 &  21.60 &    \\
14.1 &  0 &  0.04053 &  0.00020 &  723427.2 &  109154.5 &  22.20
 &  21.35 &  21.57 &    \\
68.1 &  0 &  0.04596 &  0.00017 &  723627.0 &  109503.4 &  19.13
 &  18.37 &  18.54 &    \\
60.1 &  0 &  0.04727 &  0.00023 &  723104.7 &  109479.7 &  18.61
 &  99.99 &  17.98 &    \\
\enddata
\end{deluxetable}

\clearpage

\begin{deluxetable}{cccccccccc}
\footnotesize
\tablecaption{Data for the line of sight PDCS67.\label{tbl-15}}
\tablewidth{0pt}
\tablehead{
\colhead{Num.} & \colhead{group}   & \colhead{z}
& \colhead{err-z}   & \colhead{Alpha} & \colhead{Delta}
& \colhead{V} & \colhead{I} & \colhead{R} & \colhead{type}
}
\startdata
56.1 &  0 &  0.02427 &  0.00012 &  867650.4 &  148451.0 &  18.17
 &  18.08 &  17.32 & em \\
59.1 &  0 &  0.03314 &  0.00027 &  867420.7 &  148496.2 &  99.99
 &  99.99 &  18.37 & em \\
21.1 &  0 &  0.03323 &  0.00001 &  867661.2 &  148209.8 &  19.67
 &  19.33 &  19.35 & em \\
 9.1 &  0 &  0.06550 &  0.00036 &  867704.4 &  148123.4 &  21.98
 &  20.99 &  21.26 &    \\
57.1 &  0 &  0.10789 &  0.00062 &  867347.5 &  148468.8 &  99.99
 &  99.99 &  20.81 &    \\
22.1 &  1 &  0.16664 &  0.00017 &  867132.0 &  148233.6 &  21.56
 &  20.56 &  20.98 &    \\
25.1 &  1 &  0.16773 &  0.00049 &  867353.4 &  148248.4 &  20.08
 &  19.61 &  19.66 &    \\
46.1 &  1 &  0.16852 &  0.00032 &  867537.0 &  148399.2 &  18.64
 &  17.62 &  18.07 &    \\
\enddata
\end{deluxetable}


\clearpage

\begin{thebibliography}{} 

\bibitem{}  Adami, C., Mazure, A., Katgert, P., Biviano, A., 1998 A\&A, 336,
63 a

\bibitem{}  Adami, C., Nichol, R.C., Mazure, A., et al. 1998, A\&A, 334, 765 b

\bibitem{}  Adami, C., Mazure, A., Biviano, A., Katgert, P., 1998 A\&A, 331,
493 c

\bibitem{}  Bahcall, N.A., Fan, X., \& Cen, R. 1997, 485, L53

\bibitem{}  Bertin, E., \& Arnouts, S. 1996, A$\&$AS, 117, 393

\bibitem{}  Bruzual, A.G., Charlot, S. 1993, ApJ, 405, 538

\bibitem{}  Burstein, D., Heiles, C. 1982, AJ, 110, 1507

\bibitem{}  Carlberg, R., Yee, H., Ellingson, E., et al. 1996, ApJ, 462, 32

\bibitem{}  Dussert, C., Rasigni, G., Rasigni, M., Palmari, J., Llebaria, A.
1986, Physical Review B., 34, 3528

\bibitem{}  Gunn, J.E., \& Oke, J.B. 1975, ApJ, 195, 255 

\bibitem{}  Holden, B.P., et al. 2000, AJ, in revision

\bibitem{}  Holden, B.P., Nichol, R.C., Romer, A.K., Metevier, A.,
Postman, M., Ulmer, M.P.,  \& Lubin, L.M.  1999, AJ, in press 

\bibitem{}  Holden, B.P., Romer, A.K., Nichol, R.C., Ulmer, M.P. 1997, AJ, 
114, 1701

\bibitem{}  Jones, L.R., Scharf, C., Ebeling, H., et al. 1998, ApJ, 495, 100

\bibitem{}  Katgert, P., Mazure, A., Perea., J., et al. 1996, A\&A, 310, 8

\bibitem{}  Kurtz, M.J., \& Mink, D.J. 1998, PASP, 110, 934

\bibitem{}  Landolt, A.U. 1992, AJ, 104,340  

\bibitem{}  Lubin, L.M., Postman, M., Oke, J.B., et al. 1998, AJ, 116, 584

\bibitem{}  Lubin, L.M. 1996, AJ, 112, 23

\bibitem{}  Lubin, L.M., Postman, M. 1996, AJ, 111, 1795

\bibitem{}  Mazure, A., Katgert, P., den Hartog, R., et al. 1996, A\&A, 310,
31

\bibitem{}  Nichol, R.C., Romer, A.K., Holden, B.P., et al. 1999, ApJ, 521, L21

\bibitem{}  Oke, J.B., \& Gunn, J.E. 1983, ApJ, 266, 713

\bibitem{}  Oukbir, J., \& Blanchard, A. 1997, A\&A, 317, 1

\bibitem{}  Oukbir, J., \& Blanchard, A. 1992, A\&A, 262, L21

\bibitem{}  Postman, M., Lubin, L.M., Gunn, J., et al. 1996, AJ, 111, 615

\bibitem{}  Rauzy, S., Adami, C., Mazure, A. 1998, A$\&$A, 337, 31

\bibitem{}  Reichart, D.E., Nichol, R.C., Castander, F.J., et al. 1999, ApJ,
 518, 521

\bibitem{}  Romer, A.K., Nichol, R.C., Holden, B.P., et al. 2000, ApJS, 
in press

\bibitem{}  Rosati, P., Della Cecca, R., Norman, C., Giacconi, R. 1998,
ApJ, 492, L21

\bibitem{}  Schlegel, D.J., Finkbeiner, D., Davis, M. 1998, ApJ, 500, 525

\bibitem{}  Tonry, J., \& Davis, M. 1979, AJ, 84, 1511

\bibitem{}  Vikhlinin, A., McNamara, B.R., Forman, W., Jones, C. 1998,
ApJ, 502, 558

\bibitem{}  Yee, H.K.C., Ellingson, E., Carlberg, R.G. 1996, ApJS, 102, 269

\end{thebibliography}
\end{document}